\documentclass[twocolumn,english]{revtex4-2}
\usepackage[T1]{fontenc}
\usepackage[latin9]{inputenc}
\usepackage[a4paper]{geometry}
\usepackage{hyperref}
\usepackage{amssymb}
\usepackage{amsmath,amssymb,amsfonts}
\usepackage{comment}
\usepackage{environ}
\usepackage{braket}
\usepackage[normalem]{ulem}
\usepackage{tikz-feynman}
\usepackage{float}
\usepackage[dvipsnames]{xcolor}
\newcommand{\symbex}{\tikz{
\draw (-3pt, 2pt) -- (-2pt, 0);
\draw (-3pt, -2pt) -- (-2pt, 0);
\draw (-2pt, 0) -- (2pt, 0);
\draw (3pt, 2pt) -- (2pt, 0);
\draw (3pt, -2pt) -- (2pt, 0);
    }}
\usetikzlibrary{shapes.misc}
\usetikzlibrary{decorations.markings}
\tikzset{
  midarrow/.style={
    postaction={decorate},
    decoration={markings, mark=at position .5 with {\arrow{>}}}
  }
}

\def\be{\begin{equation}}
\def\ee{\end{equation}}
\def\p{\partial}

\newcommand{\nno}{\nonumber}

\def\e{\epsilon}

\def \lin {\langle{\text{in}}|}
\def \rin {|{\text{in}}\rangle}
\newcommand{\disc}{\small\text{disc}}

\newcommand{\intinf}{\int_{-\infty}^{\infty}}
\newcommand{\intsinf}{\int_{0}^{\infty}}

\NewEnviron{eqn}{
\begin{align}
\begin{split}
  \BODY
\end{split}
\end{align}
}

\NewEnviron{eqn*}{
\begin{align*}
\begin{split}
  \BODY
\end{split}
\end{align*}
}

\begin{document}
\widetext
\title{Cosmological Correlator Discontinuities from Scattering Amplitudes}
\author{Chandramouli Chowdhury$^{a}$}
\author{Sadra Jazayeri$^{b}$}
\author{Arthur Lipstein$^{c}$}
\author{Joe Marshall$^{c}$}
\author{Jiajie Mei$^{d}$}
\author{Ivo Sachs$^{e}$}
\affiliation{$^{a}$School of Mathematical Sciences, University of Southampton, Southampton SO17 1BJ, UK}
\affiliation{$^{b}$Abdus Salam Centre for Theoretical Physics, Imperial College, London, SW7 2AZ, UK}
\affiliation{$^{c}$Department of Mathematical Sciences, Durham University, Durham, DH1 3LE, UK}
\affiliation{$^{d}$Institute of Physics, University of Amsterdam, Amsterdam, 1098 XH, The Netherlands}
\affiliation{$^{e}$Arnold-Sommerfeld-Center for Theoretical Physics, Ludwig-Maximilians-Universit\"at M\"unchen, Theresienstr. 37, D-80333 Munich, Germany}

\begin{abstract}

Recent theoretical work has revealed that basic observables of quantum field theory in de Sitter space, known as in-in or cosmological correlators, exhibit surprisingly simple mathematical structure reminiscent of scattering amplitudes in flat space. For many theories, this simplicity can be made manifest using a set of  ``cosmological dressing rules'' which uplift flat-space Feynman diagrams to in-in correlators in de Sitter space by attaching auxiliary propagators to the interaction vertices. In this paper, we show that discontinuities of cosmological correlators with respect to internal energy variables can be obtained by applying auxiliary propagators to unitarity cuts of flat space Feynman diagrams. Moreover, discontinuities with respect to external energy variables can be obtained by cutting auxiliary propagators attached to Feynman diagrams. This observation in turn implies highly non-trivial constraints on cosmological correlators in the form of simple sum rules. We illustrate these ideas in a number of examples at tree-level and 1-loop for conformally coupled scalar theories, although they hold more generally. Finally, we show how to reconstruct cosmological correlators from their discontinuities using dispersion relations, providing a powerful new approach to computing cosmological observables by systematically reconstructing them from data uplifted from flat space.        
\end{abstract}

\maketitle

\section{Introduction}\label{sec:intro}

According to the paradigm of inflation, the early universe underwent a period of accelerated expansion and was approximately described by a geometry known as four-dimensional de Sitter space (dS$_4$) \cite{Starobinsky:1980te,Mukhanov:1981xt,Guth:1980zm,Linde:1981mu}. The basic observables in this background are in-in or cosmological correlators which are traditionally computed using the Schwinger-Keldysh formalism \cite{Weinberg:2005vy} or by squaring a quantity known as the cosmological wavefunction and performing a path integral over boundary values of bulk fields \cite{Maldacena:2002vr}. In recent years there has been considerable effort to develop methods to compute wavefunction coefficients inspired by the study of scattering amplitudes and conformal field theory. These include methods based on symmetries \cite{Maldacena:2011nz,Bzowski:2013sza,Ghosh:2014kba, Bzowski:2019kwd,Sleight:2019mgd, Goodhew:2024eup}, factorisation \cite{Raju:2010by, Raju:2011mp, Raju:2012zr,Arkani-Hamed:2015bza,Arkani-Hamed:2018kmz,Goodhew:2020hob, Jazayeri:2021fvk, Melville:2021lst}, differential equations \cite{Arkani-Hamed:2023kig,Arkani-Hamed:2023bsv}, geometric formulations \cite{Arkani-Hamed:2017fdk,Arkani-Hamed:2024jbp, Benincasa:2024leu,  Glew:2025otn}, scattering equations \cite{Gomez:2021qfd}, and the double copy \cite{Farrow:2018yni,Armstrong:2020woi,Albayrak:2020fyp,Armstrong:2023phb, Chowdhury:2024SD}. On the other hand, recent theoretical developments have demonstrated that in many cases cosmological correlators are actually simpler than wavefunction coefficients and are intimately related to scattering amplitudes in flat space \cite{Chowdhury:2023arc,Donath:2024utn,Chowdhury:2025ohm,Glew:2025mry,Arkani-Hamed:2025mce}, although our understanding of cosmological correlators is still very primitive compared to that of scattering amplitudes. Moreover, current and upcoming cosmic microwave background and large-scale structure surveys aim to measure cosmological correlators with unprecedented precision (e.g. see \cite{Achucarro:2022qrl} and references therein). It is therefore imperative to develop new theoretical tools to compute cosmological correlators and understand their mathematical structure more deeply, as this may suggest new physical principles.

For many theories, the mathematical simplicty of cosmological correlators can be derived from Feynman diagrams in flat spacetime by applying auxiliary propagators to the vertices, collectively known as the 
``cosmological dressing rules''. This was first derived for conformally coupled and massless scalar theories in dS$_4$ \cite{Chowdhury:2025ohm} building on earlier work which recast in-in correlators in terms of a shadow formalism in Euclidean Anti-de Sitter space (EAdS) \cite{Sleight:2020obc,Sleight:2021plv,DiPietro:2021sjt,Heckelbacher:2022hbq}, and these developments have recently been extended to spinning theories in \cite{Schaub:2023scu,MdAbhishek:2025dhx,Sleight:2025dmt,Chowdhury:2025nnk}. As we will show in this paper, this formulation of cosmological correlators has an important consequence: their discontinuities can be uplifted from flat space! In more detail, we find that discontinuities with respect to energies flowing through internal propagators in dS$_4$, which we refer to as internal energies, can be directly mapped to discontinuities of flat space Feynman diagrams via the dressing rules. The latter can in turn be computed from unitarity cuts of Feynman diagrams, allowing us to directly compute discontinuities of cosmological correlators simply by dressing unitarity cuts of Feynman diagrams with auxiliary propagators. At this point one may ask what happens if we cut auxiliary  propagators attached to Feynman diagrams? This also has a beautiful answer: it simply computes the discontinuity of the the correlators with respect to energies flowing through external propagators, which we refer to as external energies. 

The ability to deduce discontinuities of cosmological observables from scattering amplitudes in flat space is conceptually profound and has several practical applications. For example, it implies a set of highly non-trivial constraints on correlators in the form of sum rules, which are not enjoyed by wavefunction coefficients. These sum rules can in principle provide useful constraints for bootstrapping correlators. Furthermore we show how to reconstruct correlators from their discontinuities using dispersion relations. Taken together, this provides a powerful new machinery for computing cosmological correlators, which we demonstrate for a few simple examples at tree-level and 1-loop in conformally coupled scalar theories, although it holds more generally as we will show in a follow-up paper \cite{long}.   

This paper is organised as follows. In section \ref{sec:review} we review in-in correlators and the dressing rules for conformally coupled theories. We then show how to map discontinuities of cosmological correlators to cuts of dressed Feynman diagrams and deduce sum rules in section \ref{sec:disccut}, and illustrate this in a number of examples in section \ref{sec:examples}. In section \ref{sec:dispersion}, we demonstrate how correlators can be bootstrapped from their discontinuities through appropriate dispersion relations.
Finally, we present our conclusions and future directions in section \ref{conclusion}. We also include a number of Appendices explaining the relation of our construction to microcausality, describing how to incorporate conformal-invariant regulator, and providing further details about the discontinuity of the 1-loop bubble integral.

\textit{Note: After this work was completed, \cite{Das:2025qsh, Colipi-Marchant:2025oin, Ansari:2026xkm} appeared which have some overlap with our results.}

\section{Review}\label{sec:review}
Let us begin by reviewing some basic definitions and concepts. We will consider in-in correlators in four-dimensional de Sitter space, which has the following metric in Poincare patch coordinates:
\begin{eqn}
    ds^2 = \frac{-d\eta^2+d \vec{x}^2}{H^2 \eta^2}.
\end{eqn}
We henceforth set the Hubble constant $H=1$. The conformal time runs $-\infty < \eta \leq \eta_0$ with the future boundary located at $\eta_0 \to 0$, on which we calculate correlation functions. The $\vec x$ coordinates span the three-dimensional spatial boundary. Since the metric depends non-trivially on time, we see that energy is not conserved. On the other hand, the translational symmetry along the boundary makes it natural to Fourier transform along boundary directions and express correlators as functions of boundary momenta. 

In this paper, we will focus on correlators of scalar field theories in dS$_4$ with polynomial interactions, which will only depend on the magnitudes of boundary momenta or their sums. In particular, we will denote the magnitude of a boundary momentum as $k_{i}=|\vec{k}_{i}|$, where $\vec{k}_{i}$ is a 3-component vector representing the boundary momentum of the $i$'th external leg of a correlator. Moreover, we will refer to sum of such variables as external energies or $x$-variables, since they appear in external propagators. Furthermore, we will denote the magnitude of sums of boundary momenta, e.g. $|\vec{k}_{i}+\vec{k}_{j}|$ as internal energies or $y$ variables, since these energies appear in internal propagators. For the remainder of the paper, we will treat the $x$ variables as being independent of the $y$ variables, which corresponds to considering off-shell correlators \cite{Cespedes:2025dnq, Salcedo:2022aal}. 
In physical kinematics the $x$ and $y$ variables should be real and positive and should satisfy the Cauchy-Schwarz inequalities.

After Fourier transforming to momentum space along the boundary directions, in-in or cosmological correlators can be computed from the following expectation value:
\begin{eqn}\label{eq:CorrelatorDefn}
\braket{\phi(\vec{k}_{1})\ldots\phi(\vec{k}_{n})}=\frac{\int\mathcal{D}\phi\;\phi(\vec{k}_{1})\ldots\phi(\vec{k}_{n})\left|\Psi[\phi]\right|^{2}}{\int\mathcal{D}\phi\left|\Psi[\phi]\right|^{2}},
\end{eqn}
where we restrict to a scalar theory in the bulk for simplicity, $\phi$ denotes the boundary value of this field which must be integrated over, and $\Psi(\phi)$ is the cosmological wavefunction \cite{Hartle:1983ai}. The latter admits a perturbative expansion in terms of wavefunction coefficients which can be obtained by Wick rotation of Witten diagrams in EAdS \cite{Maldacena:2002vr,McFadden:2009fg}.

As mentioned in the introduction, cosmological correlators often exhibit surprising mathematical simplicity which can be made manifest by dressing flat space Feynman diagrams with certain auxiliary propagators which uplift them to de Sitter space \cite{Chowdhury:2025ohm}. We will now explain how this works for conformally coupled scalars (which have mass $m^2=2$) with $\phi^4$ and $\phi^3$ interactions in dS$_4$, although similar dressing rules can also be defined for massless scalar theories and spinning theories \cite{Chowdhury:2025nnk}. In previous references, the ``cosmological dressing rules'' were derived in Euclidean signature, although it is straightforward to adapt them to Lorentzian signature by introducing the usual Feynman $i \epsilon$ prescription, where the sign of $\epsilon$ is chosen in the usual way to ensure that the poles do not collide with the integration contour after Wick rotating back to Euclidean signature. 

The first step is to start with a flat space Feynman diagram, relax energy conservation, and attach an auxiliary one-dimensional propagator to each interaction vertex whose energy is equal to the sum of the energies of all the other internal lines attached to that vertex, which we shall denote as $p_{\rm{tot}}$. The next step is to attach the other end of each auxiliary propagator to a common point and impose energy conservation at that point. We then end up with a dressed diagram of the form  
\begin{eqn}
\label{dressingexample}
\centering
\begin{tikzpicture}[baseline]
    \draw (-1.25,0.25) -- (-1,0);
    \draw (-1.25,-0.25) -- (-1,0);
    \draw (-1.3,0) -- (-1,0);
    \draw (-1,0) --(0,0) --(1,0) --(2,0);
    \draw (0.25,-0.25) -- (0,0);
    \draw (-0.25,-0.25) -- (0,0);
    \draw (1.25,-0.25) -- (1,0);
    \draw (0.75,-0.25) -- (1,0);
    \draw (2.25,0.25) -- (2,0);
    \draw (2.25,-0.25) -- (2,0);
    \draw (2.3,0) -- (2,0);
    \draw [dashed] (-1,0) --(0.5,1) -- (2,0);
    \draw [dashed] (0,0) --(0.5,1) --(1,0);
\end{tikzpicture}.
\end{eqn}
Finally, integrate over all the independent energies running through auxiliary propagators. The precise form of the auxiliary propagators is theory-dependent. For the conformally coupled $\phi^4$ theory, they are given by
\begin{eqn}
    \begin{tikzpicture}[baseline]
        \draw (-1,0) -- (0,0);
        \draw (0,0) -- (1,0);
        \draw[dashed] (0,1) -- (0,0);
        \node at (-1,0.2) {\(x_{\rm ext}\)};
        \node at (0.5,0.4) {\(p_{\rm tot}\)};
    \end{tikzpicture} = \frac{x_{\rm ext}}{p_{\rm{tot}}^2-x_{\rm ext}^2 + i \e},
    \label{eq:phi4Auxprop}
\end{eqn}
where the auxiliary propagator is represented by a dashed line, the solid line labelled by $p_{\rm tot}$ collectively represents the internal legs which are attached to the vertex, and the solid line labelled by $x_{\rm ext}$ denotes the external legs attached to the vertex, with $x_{\rm ext}$ denoting the sum of external energies at the vertex. In practice, there will be four solid lines attached to each vertex, but here we have only depicted two because we are not specifying which of the four legs are internal or external. Note that the energy flowing through the auxiliary propagator is also given by $p_{\rm tot}$ and putting the auxiliary propagator on-shell sets $p_{{\rm tot}}=x_{{\rm ext}}$, restoring energy conservation. Moreover, if there are no external legs attached to the vertex then $x_{\rm{ext}}\rightarrow 0$ and we obtain a $\delta\left(p_{{\rm tot}}\right)$, so no energy flows through the auxiliary propagator. 

In conformally coupled $\phi^4$ theory, a dressed diagram with $n$ external legs and $v$ vertices with at least one external leg has the general form 
\begin{eqn}\label{corrdefnphi4}
C_n(\{x\}, \{\vec y\}) =  \intinf \prod_{i = 1}^{v} \frac{dp_i \delta(\sum_j p_{j}) }{p_i^2 - x_i^2 + i \e} A_n(\{p\}; \{\vec y\})
\end{eqn}
where $A_n$ is a flat space Feynman diagram in massless $\phi^4$ theory which depends on Lorentz-invariant combinations formed from the 4-vectors of the form $(p, \vec y)$. 
Note that the delta function in the $p$ variables comes from the vertex where all the auxiliary propagators meet. The in-in correlator is then obtained by multiplying by $x_1 \cdots x_v$ and summing over all dressed Feynman diagrams. We must also include an overall factor of $\eta_{0}^{n}\left(k_{1}\cdots k_{n}\right)^{-1}\delta^{3}\left(\sum_{i=1}^{n}\vec{k}_{i}\right)$, although we will drop this factor in practice.

For conformally coupled $\phi^3$ theory, there are two types of auxiliary propagators:
\begin{eqn}
    \begin{tikzpicture}[baseline]
        \draw (-1,0) -- (0,0);
        \draw (0,0) -- (1,0);
        \draw[dashed] (0,1) -- (0,0);
        \node at (-1,0.2) {\(x_{\rm ext}\)};
        \node at (0.5,0.4) {\(p_{\rm tot}\)};
    \end{tikzpicture} &= \int_{x_{\rm ext}}^\infty ds \, \frac{p_{\rm tot}}{p_{\rm tot}^2-s^2 + i \e}\\
    \begin{tikzpicture}[baseline]
        \draw (-1,0) -- (0,0);
        \draw (0,0) -- (1,0);
        \draw[dotted] (0,1) -- (0,0);
        \node at (-1,0.2) {\(x_{\rm ext}\)};
        \node at (0.5,0.4) {\(p_{\rm tot}\)};
    \end{tikzpicture} &=\pi.
    \label{eq:phi3Auxprop}
\end{eqn}
The first type must occur an even number of times for any given diagram while the second type can occur an arbitrary number of times. To obtain an in-in correlator, we must sum over all possible ways of attaching these auxiliary propagators to each flat space Feynman diagram subject to these restrictions. For the purpose of illustration, we will restrict our attention to diagrams which are only dressed with the first type of auxiliary propagator (which must have an even number of external legs). They take the form
\begin{align}\label{corrdefnphi3}
D_n(\{x\}, \{\vec y\}) &=  \bigg(\prod_{i = 1}^{v} \int_{x_i}^\infty ds_i \intinf dp_i   \frac{p_i  }{p_i^2 - s_i^2 + i \e}\bigg)\nno \\
&\times\delta(\sum_j p_{j}) A_n(\{p\}; \{\vec y\}).
\end{align} 
It is straightforward to extend our results to diagrams which also contain the other type of auxiliary propagator since it has trivial kinematic dependence.

\section{From discontinuities to cuts\label{sec:disccut}}
In this section, we will use the dressing rules to analyse discontinuities of cosmological correlators in conformally coupled $\phi^4$ and $\phi^3$ theory. In more detail, we will show that discontinuities with respect to internal energies, or $y$ variables, can be mapped to those of flat space Feynman diagrams, which can in turn be computed from unitarity cuts using well-known methods. Moreover, discontinuities with respect to external energies, or $x$ variables, arise from cutting auxiliary propagators. Hence, all discontinuities of cosmological correlators can be directly lifted from flat space scattering amplitudes, either by dressing flat space unitarity cuts with auxiliary propagators or by dressing flat space diagrams with cut auxiliary propagators. 

The discontinuity of a function $f(z)$ across the real line is defined as (e.g. \cite{Britto:2024mna})
\begin{eqn}\label{discdefn}
\disc_{z} f(z) \equiv \frac{1}{2\pi i} \lim_{\e \to 0^+} f(z + i \e) - f(z - i \e). 
\end{eqn}
By using \eqref{discdefn} we shall relate these discontinuities to cuts of propagators, which is facilitated by the identity 
\begin{eqn}\label{discid}
\disc_{z}\frac{1}{s- z} = \frac{1}{2\pi i}\lim_{\e \to 0^+}\frac{\e}{(s-z)^2+\e^2}=\delta(s - z),
\end{eqn}
for real $z$. When doing so, we may set the Feynman $i \epsilon$ in those propagators to zero. Let us now show how this works for general discontinuities with respect to $x$ and $y$ variables. We will present numerous concrete examples in section \ref{sec:examples}.

\subsection*{y discontinuities}
Let us first discuss the discontinuity with respect to variables of the type $y^2$. 
From the dressing rules in \eqref{corrdefnphi4} and \eqref{corrdefnphi3}, it is clear that a discontinuity with respect to $y^2$ will not affect the auxiliary propagators (as the variables $x, y$ are assumed to be independent) and only act on the flat space Feynman diagram $A_n$. For example, the discontinuity with respect to $y_1^2 \equiv\vec y_1^2$ in \eqref{corrdefnphi4} gives
\begin{eqn}
\disc_{y_1^2}C_n = \bigg(\int \prod_{i = 1}^{v}  \frac{ dp_i \delta(\sum_j p_{j})}{p_i^2 - x_i^2 + i \e}\bigg)  \disc_{y_1^2}A_n.
\end{eqn}
$A_n$ must be function of Lorentz-invariant quantities of the form $s=-p^2+y^2$.
Since these variables are linearly related to $y^2$, \eqref{discdefn} implies that $\disc_{y^2}$ of a cosmological correlator can be computed from $\disc_{s}$ of the flat space Feynman diagram. The latter can in turn be computed using unitarity cuts (see for example \cite{Britto:2024mna} for a modern review of flat-space cutting rules) which are implemented by replacing cut propagators with $\delta^+(p^2-y^2) \equiv \Theta(p) \delta(p^2-y^2)$. Note that the kinematical region for which the flat space discontinuities are non-vanishing in turn determines the domain of the subsequent $p$-integrations. For instance, for a 1-loop bubble diagram $s<0$ implies $p^2>y^2$. 

In summary, discontinuities with respect to $y^2$ variables can be obtained by applying the dressing rules to flat space cuts. In an upcoming paper \cite{long}, we describe how to derive this more systematically from the largest time equation \cite{Veltman:1963th, tHooft:1973wag}. 
We remark that $y^2$ discontinuities can also be derived from microcausality, i.e. the requirement that local operators must commute at spatial separations \cite{Tong:2021wai, Ema:2024hkj, Hui:2025aja}. In Appendix \ref{sec:causality} we demonstrate this for 2-site graphs. We will discuss the implications of causality for higher-site graphs, such as triangles, in \cite{long}.

\subsection*{x discontinuities}
Next, let us consider discontinuities with respect to $x^2$ variables.
As an example, let us compute 
\begin{eqn}\label{singlexdiscex1}
&\disc_{x_1^2} C_n =  \int \prod_{i = 2}^v \frac{ dp_i  dp_1 \delta(p_1^2-x_1^2)}{p_i^2 - x_i^2 + i \e} \delta(\sum p_j)A_n(\{p\}, \{\vec y\}),
\end{eqn}
where we noted that taking the discontinuity simply replaces $(p_1^2-x_1^2)^{-1}$ by a delta function.
Discontinuities of $D_n$ in \eqref{corrdefnphi3} can be computed in a similar manner except that the $s$-integral in \eqref{eq:phi3Auxprop} gives rise to $\Theta$-functions:
\begin{align}\label{eq:xCutDefn}
&\disc_{x^2} \int_0^\infty \frac{p \,ds}{p^2 - (s+x)^2 } = \int_0^\infty ds \, p \, \delta(p^2 - (s+x)^2)\,.\nonumber \\ &=   \frac{1}{2}\Theta(p - x)-\frac{1}{2}\Theta(-p - x),
\end{align}
where we assume $x> 0$. The presence of $\Theta-$functions is not surprising as the auxiliary propagators in $D_n$ take a similar form to those in $C_n$ after differentiating with respect to $x$. In summary, we find that discontinuities with respect to $x^2$ variables correspond to replacing $(p^2-x^2)^{-1}$ terms appearing in auxiliary propagators with delta functions. In practice, we can actually replace such terms with $\delta^+(p^2-x^2)$, as we will see in explicit examples in the next section and prove more generally in \cite{long}. This procedure can be generalized to compute sequential discontinuities of both $x$, $y$ variables, as we will illustrate for the 1-loop triangle correlator in the next section.

\subsection*{Flat space limit and sum rules}
Since the auxiliary propagators promote flat space diagrams to cosmological correlators, it is expected that we should recover the former by cutting all auxiliary propagators. Indeed, after setting all the auxiliary propagators on-shell in conformally coupled $\phi^4$ theory, we obtain the flat-space amplitude including an energy-conserving delta function:
\begin{eqn}\label{flatspacelimit}
&\disc_{x_{v}^2} \cdots \disc_{x_1^2} C_n \propto  
\delta(E) A_n(\{x\}; \{\vec y\})~, 
\end{eqn}
where $E=k_1 + \cdots + k_n$ is the total energy and $v$ denotes the number of external energy variables. 
This gives a new way to compute the flat-space limit for the correlator using sequential discontinuities. 
By keeping track of higher-order terms in \eqref{discdefn} one can systematically study corrections to the flat space limit of the correlator which were also studied in \cite{Arkani-Hamed:2025mce}.

In the presence of UV-divergences there are additional contributions when using a dS-invariant regularization scheme. Indeed, it is known \cite{Raju:2012zr, Chowdhury:2023arc} that the flat space limit of renormalized, dS-invariant correlators also involves contributions from branch cuts in the total energy. We discuss an example of this in appendix \ref{app:dSinv}. 
In a similar way, cutting only a subset of auxiliary propagators yields products of lower-point correlators and amplitudes, which are referred to as partial energy singularities \cite{long}. For $D_n$, cutting all the auxiliary propagators does not directly give an energy conserving delta function but we obtain the delta function after acting further with a derivative with respect to each $x$ variable. We will see an explicit example of this in the next section.

From equation \eqref{flatspacelimit} we find that performing sequential discontinuities with respect to all $x^2$ variables must vanish when energy is not conserved. Noting that the discontinuity with respect to an $x^2$ is equivalent to summing over $\pm \sqrt{x^2}\equiv \pm x$, we find that
\begin{eqn}\label{sumrule}
\sum_{\{s\} = \pm 1} \Bigl(\prod_i s_i\Bigr) C_n(s_1 x_1, s_2 x_2, \cdots, s_v x_v; \{ \vec y\}) = 0.
\end{eqn}
This provides interesting non-trivial constraints on cosmological correlators  (see also \cite{Donath:2024utn, Werth:2024mjg}). As we will see in the next section, a similar sum rule is obeyed by a $\phi^3$ correlators.
This provides powerful new  consistency relations on cosmological correlators which are not in general enjoyed by wavefunction coefficients.

\section{Examples of cutting rules}\label{sec:examples}
Let us now illustrate the results of the previous section with a number of concrete examples at tree-level and 1-loop.

\subsection*{Tree-level exchange}
The tree-level 4-point correlator in conformally coupled $\phi^3$ theory can be constructed by dressing massless exchange diagrams in flat space with the auxiliary propagators in \eqref{eq:phi3Auxprop}. For concreteness, we will focus our attention on the s-channel contribution to this correlator obtained by dressing with dashed propagators since this gives the most non-trivial part of the correlator:
\begin{align}
\label{dressforD2}
D_4^{\symbex} &= \begin{tikzpicture}[baseline]
\draw (-0.75, 0.5) -- (-0.5, 0);
\draw (-0.75, -0.5) -- (-0.5, 0);
\draw (0.75, 0.5) -- (0.5, 0);
\draw (0.75, -0.5) -- (0.5, 0);
\draw (-0.5,0) -- (0.5,0);
\draw[dashed] (0.5, 0) -- (0, -0.75);
\draw[dashed] (0, -0.75) -- (-0.5, 0);
\node at (-1, 0) {$x_1$};
\node at (1, 0) {$x_2$};
\node at (0, 0.35) {$ \vec{y}$};
\end{tikzpicture}  \\
&=\int_{x_1}^{\infty} ds_1 \int_{x_2}^{\infty} ds_2\int\frac{p^2 dp}{(p^2 - s_1^2 ) (p^2 - s_2^2 )(p^2 - y^2)},\nonumber
\end{align}
where $x_1= k_1+k_2$, $x_2=k_3+k_4$, $y^2=(\vec{k}_1+\vec{k}_2)^2$, and we suppress $i \e$'s for compactness. The result of the integrals is given by dilogarithms \cite{Arkani-Hamed:2015bza}:
\begin{eqn}
\label{D2explicit}
    D_4^{\symbex} = \frac{1}{2 y} &\bigg( \text{Li}_2\big( \frac{x_1-y}{x_1+x_2}\big)+\text{Li}_2\big( \frac{x_2-y}{x_1+x_2}\big)\\
    &+\log\big(\frac{x_1+y}{x_1+x_2}\big) \log\big(\frac{x_2+y}{x_1+x_2}\big)-\frac{\pi^2}{6}\bigg)\,.
\end{eqn}
If we compute the discontinuity with respect to $y^2$ according to \eqref{discdefn}, we see that this effectively replaces the flat space Feynman propagator $(p^2-y^2)^{-1}$ with a delta function: 
\begin{eqn}
\label{D2disc}
\disc_{y^2}D_4^{\symbex}=&\int\limits_{x_1}^{\infty} ds_1 \int\limits_{x_2}^{\infty} ds_2\int\frac{p^2 dp \;\delta^+(p^2 - y^2)}{(p^2 - s_1^2 ) (p^2 - s_2^2 )}\\ 
=& \begin{tikzpicture}[baseline]
\draw (-0.75, 0.5) -- (-0.5, 0);
\draw (-0.75, -0.5) -- (-0.5, 0);
\draw (0.75, 0.5) -- (0.5, 0);
\draw (0.75, -0.5) -- (0.5, 0);
\draw (-0.5,0) -- (0.5,0);
\draw[dashed] (0.5, 0) -- (0, -0.75);
\draw[dashed] (-0.5, 0) -- (0, -0.75);
\node at (-1, 0) {$x_1$};
\node at (1, 0) {$x_2$};
\node at (0, 0.2) {$\vec{y}$};
\draw (-0.25,-0.3) [red]-- (0.25,0.3);
\end{tikzpicture},
\end{eqn}
where the red slash denotes a cut. Note that we can replace the delta function with $\delta^+(p^2-y^2)$ using the $p\rightarrow -p$ symmetry of the integrand. Hence, we see that the discontinuity with respect to $y^2$ is simply obtained by dressing a flat space unitarity cut. 
Performing the integral in \eqref{D2disc} gives 
\begin{eqn}
\disc_{y^2}D_4^{\symbex}=-\frac{
    \log \frac{x_1 + y}{x_1 - y} \log \frac{x_2 + y}{x_2 -y}
    }{2y}.
    \label{bubmaxdisc}
\end{eqn}
Using the explicit formula in \eqref{D2explicit}, we find that the discontinuity with respect to $y^2$ is given by $D_4^{\symbex}(-y)-D^{\symbex}_4(y)$ in the kinematic regime $x_1, x_2 > y > 0.$ Next, let us consider what happens when we compute discontinuities with respect to external energy variables. Following the discussion of $x$ discontinuities in section \ref{sec:disccut} we find that 
\begin{eqn}
\disc_{x_1^2} \disc_{x_2^2}D_4^{\symbex}&=\int \frac{ \Theta(p-x_1) \Theta(p-x_2) dp }{(p^2-y^2+i\e)}\\
&=
    \begin{tikzpicture}[baseline]
\draw (-0.75, 0.5) -- (-0.5, 0);
\draw (-0.75, -0.5) -- (-0.5, 0);
\draw (0.75, 0.5) -- (0.5, 0);
\draw (0.75, -0.5) -- (0.5, 0);
\draw (-0.5,0) -- (0.5,0);
\draw[dashed] (0.5, 0) -- (0, -0.75);
\draw[dashed] (-0.5, 0) -- (0, -0.75);
\node at (-1, 0) {$x_1$};
\node at (1, 0) {$x_2$};
\node at (0, 0.2) {$\vec{y}$};
\draw (-0.4,-0.4) [red]-- (0.4,-0.4);
\end{tikzpicture}~,
\end{eqn}
where we have used symmetry of the integrand to replace $(p^2-x^2)^{-1}$ terms in the auxiliary propagators with $\delta^+(p^2-x^2)$, as discussed below \eqref{eq:xCutDefn} (this is what the red slash indicates). 
This integral can be evaluated in terms of logarithms.
Hence, we see that discontinuities with respect to external energies are obtained by cutting auxiliary propagators attached to flat space Feynman diagrams. In fact, the flat space Feynman diagram is recovered by the taking derivatives $\p_{x_1} \p_{x_2}$, yielding
\begin{eqn}\label{px1px2discx1discx2}
    \p_{x_1} \p_{x_2}\disc_{x_1^2} \disc_{x_2^2}D_4^{\symbex} 
    \propto \frac{\delta(x_1+x_2)}{-x_1^2+y^2}.
\end{eqn}
The appearance of the derivatives is not surprising since a $\phi^3$ interaction in dS$_4$ maps to a time-dependent interaction in flat space after a Weyl transformation. Furthermore, it can be verified that the sum rule \eqref{sumrule} is also satisfied exactly in this particular case. More generally, the cut of all auxiliary propagators can give at most a sum of contact terms. This will be discussed in greater detail in \cite{long}.

\subsection*{1-loop bubble}

Next, let us consider the 1-loop 4-point in-in correlator in conformally coupled $\phi^4$ theory, which can be obtained by summing over the three channels in flat space dressed with the auxiliary propagators in \eqref{corrdefnphi4}. In particular, the s-channel contribution is given by 
\begin{eqn}\label{eq:bub}
C_4^{\circ} &= \begin{tikzpicture}[baseline, shift = {(0, 0.25)}]
\draw (-1, 0.5) -- (-0.5, 0);
\draw (-1, -0.5) -- (-0.5, 0);
\draw (1, 0.5) -- (0.5, 0);
\draw (1, -0.5) -- (0.5, 0);
\draw (0, 0) circle (0.5);
\node at (-0.8 ,0) {$x_1$};
\node at (0.8 ,0) {$x_2$};
\draw[dashed] (0.5, 0) -- (0, -1);
\draw[dashed] (0, -1) -- (-0.5, 0);
\end{tikzpicture}
\\
&\!=\! \int \frac{ dp}{(p^2 - x_1^2)(p^2 - x_2^2)} \int\frac{d^4L}{L^2 (L + K)^2},
\end{eqn}
where $K_\mu = (p, \vec y)$ and $x_1,x_2,$ and $y$ are the same as in the previous example. The integral over loop momentum is UV divergent and needs to be regulated. Using a cutoff on the comoving momenta one obtains
\begin{align}\label{bubhardcutoff}
C_4^{\circ} &= \frac{1}{4x_1 x_2(x_1 + x_2)} \\ 
&\times \Bigg[ 
\log\frac{(x_1 + y)(x_2 + y)}{4\Lambda^2} + \frac{x_1 + x_2}{x_1 - x_2} \log\frac{x_2 + y}{x_1 + y} \Bigg]\nno
\end{align}
in the physical regime with $x_i>y>0$. Note that this regulator breaks conformal symmetry corresponding to the isometries of dS$_4$ (we will discuss a dS-invariant regulator in Appendix \ref{app:dSinv}). 

Now let's compute some discontinuities. First, consider the discontinuity with respect to $x_1^2$. Following the discussion in section \ref{sec:disccut}, this replaces the cut auxiliary propagator with $\delta^+(p^2-x_1^2)$ yielding
\begin{eqn}\label{bubxcut}
\disc_{x_1^2} C_4^{\circ} = \begin{tikzpicture}[baseline, shift = {(0, 0.25)}]
\draw (-1, 0.5) -- (-0.5, 0);
\draw (-1, -0.5) -- (-0.5, 0);
\draw (1, 0.5) -- (0.5, 0);
\draw (1, -0.5) -- (0.5, 0);
\draw (0, 0) circle (0.5);
\draw[dashed] (0.5, 0) -- (0, -1);
\draw[dashed] (-0.5, 0) -- (0, -1);
\draw[thick] (-0.5,-1) [red]-- (0,-0.65);
\end{tikzpicture} = \frac{x_1^{-1}}{x_1^2 - x_2^2} \log\frac{x_1^2 - y^2}{\Lambda^2}.
\end{eqn}
It is easy to check using \eqref{bubhardcutoff} that this is equal to $C_4^{\circ}(x_1, x_2; y) - C_4^{\circ}(-x_1, x_2; y)$. 

Next, let us consider the discontinuity with respect to $y^2$. Following the discussion in section \ref{sec:disccut}, since $y^2$ only appears in the flat space part of the dressed diagram, this can be computed by dressing the unitarity cut of the 1-loop bubble diagram in flat space for which there is a well-known expression \cite{Peskin:1995ev}: 
\begin{align}\label{eq:4pt-ycut}
&\disc_{y^2}C_4^{\circ} 
= \begin{tikzpicture}[baseline, shift = {(0, 0.25)}]
\draw (-1, 0.5) -- (-0.5, 0);
\draw (-1, -0.5) -- (-0.5, 0);
\draw (1, 0.5) -- (0.5, 0);
\draw (1, -0.5) -- (0.5, 0);
\draw (0, 0) circle (0.5);
\draw[dashed] (0.5, 0) -- (0, -1);
\draw[dashed] (-0.5, 0) -- (0, -1);
\draw[thick] (0,-0.75) [red]-- (0,0.75);
\end{tikzpicture}  \\
&=\intinf \frac{dp}{(p^2 - x_1^2) (p^2 - x_2^2)} \int d^4 L \delta^+(L^2) \delta^+((L + K)^2). \nno
\end{align}
This integral is UV finite and yields the following:
\begin{align}\label{bubmaxdiscn}
&\disc_{y^2} C_4^{\circ} = \frac{x_1 \log\big(\frac{-x_2 + y+i\epsilon}{y +x_2} \big)-x_2 \log\big(\frac{-x_1 + y+i\epsilon}{y + x_1} \big)}{2x_1 x_2(x_1^2-x_2^2)}\\
=& \frac{x_1 \log\big(\frac{x_2 - y-i\epsilon}{y +x_2} \big)-x_2 \log\big(\frac{x_1 - y-i\epsilon}{y + x_1} \big)}{2x_1 x_2(x_1^2-x_2^2)}+\frac{\pi i}{x_1x_2(x_1+x_2)}\,.\nonumber
\end{align}
This is in agreement with the $y^2$-discontinuity of the integral \eqref{eq:bub} evaluated in complex kinematics, as we show in in Appendix \ref{sec:Lbub}. Furthermore, using \eqref{bubhardcutoff} we find that \eqref{bubmaxdiscn}  equals 
$C_4^\circ(y) - C_4^\circ(-y)$ , up to the additional term $\frac{i \pi}{x_1+x_2}$, which encodes the discontinuity of the flat space Feynman diagram which arises from the residue of the $E=x_1+x_2$ pole. 
Note that even though the maximal cut of the diagram in flat space does not contain a logarithm, we obtain one after uplifting to dS$_4$ due to the $p$ integral, which computes the discontinuity across the square root branch cut in \eqref{bubhardcutoff}.

Let us now demonstrate how the sum rule in \eqref{sumrule} is realised. After integrating out the auxiliary propagator energies and the energy component of the loop momentum in \eqref{eq:bub}, we are left with the following 3d loop integral:
\begin{align}
C_4^{\circ}&=\int d^{3}l_1\,B(x_{1},x_{2},l_1, l_2);\\
B(x_{1},x_{2},l_1, l_2)&=\frac{l_{12}+x_{1}+x_{2}}{x_1 x_2 \left(x_{1}+x_{2}\right)l_{1}l_{2}\left(l_{12}+x_{1}\right)\left(l_{12}+x_{2}\right)}.\nno
\end{align}
Here $\vec{l}_{2}=\vec{l}_1+\vec{y}$, and $l_{12}=|\vec{l}_1|+|\vec{l}_2|$. It is easy to see that the loop integrand obeys the following four-term relation:
\begin{equation}
\sum_{s_i=\{1,-1\}} s_1 s_2
B\left(s_1 x_{1},s_2 x_{2},l_1,l_2\right)=0.
\end{equation}
By extension, we see that the integrated expression in \eqref{bubhardcutoff} must also obey this four term sum rule, which is precisely the statement of \eqref{sumrule}.

\subsection*{1-loop triangle}

As our final example, let's consider the 1-loop 6-point correlator in $\phi^4$ theory, which is obtained by dressing flat space triangle diagrams with the auxiliary propagators in \eqref{eq:phi4Auxprop}. Let us focus on a single dressed diagram contributing to the correlator:
\begin{align}
\label{triangle}
&C_6^\Delta = \begin{tikzpicture}[baseline, shift = {(0, 0.1)}, scale=0.5]
\coordinate (A) at (-1, -0.5);
\coordinate (B) at (1, -0.5);
\coordinate (C) at (0, 1);
\coordinate (N) at (1.5, 0.7);
\coordinate (Aleft) at (-2, -0.5);   
\coordinate (Adown) at (-1.5, -1.25); 
\coordinate (Bright) at (2, -0.5);    
\coordinate (Bdown) at (1.5, -1.25);  
\coordinate (Cright) at (0.5, 1.866); 
\coordinate (Cleft) at (-0.5, 1.866); 
\draw (A) -- (B);
\draw (B) -- (C);
\draw (C) -- (A);
\draw (A) -- (Aleft);   
\draw (A) -- (Adown);   
\draw (B) -- (Bright);  
\draw (B) -- (Bdown);   
\draw (C) -- (Cright);  
\draw (C) -- (Cleft);   
\draw[dashed] (N) -- (A);
\draw[dashed] (N) -- (B); 
\draw[dashed] (N) -- (C);
\node at (-1 -0.35, -0.25) {$x_2$};
\node at (1 -0.25, -0.5-0.35) {$x_3$};
\node at (0 -0.45,1) {$x_1$};
\node at (0,2) {$\vec y_1$};
\node at (-1 -0.95, -0.5 -0.5) {$\vec y_2$};
\node at (1 +1 , -0.5-0.5) {$\vec y_3$};
\end{tikzpicture}  \\
&=  \int_{-\infty}^{\infty} dp_1 dp_2 \frac{1}{(p_1^2 - x_1^2) (p_2^2 - x_2^2) \big( (p_1 + p_2)^2 - x_3^2 \big)} A_6, \nonumber
\end{align} 
where $x_1= k_{12}$, $x_2 = k_{34}$, $x_3 = k_{56}$, and $A_6$ is a flat space triangle diagram:
\begin{equation}
A_6 
=\int_0^{\infty} d^4 L\frac{1}{L^2  (L + P_1)^2 (L + P_2)^2},
\end{equation}
where $P_1^{\mu} = (p_1, \vec y_1)$, $P_2^{\mu}= (p_2, \vec y_2)$, with $\vec y_1 = \vec k_1 + \vec k_2$, $\vec y_2 = \vec k_{3} + \vec k_4$, and $\vec y_3 = \vec k_5 + \vec k_6$. After performing the integrals the final answer is a function of $C_6(x_1, x_2, x_3, \vec y_1, \vec y_2, \vec y_1 \cdot \vec y_2)$. 

First consider the double discontinuity with respect to $x_1^2$ and $x_2^2$. Following similar steps described in previous examples, this is implemented by cutting two auxiliary propagators:
\begin{align}\label{eq:cct}
\disc_{x_1^2} \disc_{x_2^2} C_6^\Delta &= 
\begin{tikzpicture}[baseline, shift = {(0, 0.1)}, scale=0.5]
\coordinate (A) at (-1, -0.5);
\coordinate (B) at (1, -0.5);
\coordinate (C) at (0, 1);
\coordinate (N) at (1.5, 0.7);
\coordinate (Aleft) at (-2, -0.5);   
\coordinate (Adown) at (-1.5, -1.25); 
\coordinate (Bright) at (2, -0.5);    
\coordinate (Bdown) at (1.5, -1.25);  
\coordinate (Cright) at (0.5, 1.866); 
\coordinate (Cleft) at (-0.5, 1.866); 
\draw (A) -- (B);
\draw (B) -- (C);
\draw (C) -- (A);
\draw (A) -- (Aleft);   
\draw (A) -- (Adown);   
\draw (B) -- (Bright);  
\draw (B) -- (Bdown);   
\draw (C) -- (Cright);  
\draw (C) -- (Cleft);   
\draw[dashed] (N) -- (A);
\draw[dashed] (N) -- (B); 
\draw[dashed] (N) -- (C);
\draw (1,1.2) [red]-- (1,0.1);
\end{tikzpicture} \\
&=\frac{1}{x_1 x_2\left(x_3^2-(x_1+x_2)^2 \right)}A_6(\vec{y}_1, \vec{y}_2, x_1, x_2). \nonumber 
\end{align}
Note that we obtain the flat space amplitude without energy conservation. At the same time, the double discontinuity can be expressed by summing over sign flips of $x_1$ and $x_2$ following the discussion around \eqref{sumrule}:
\begin{equation}
\disc_{x_1^2} \disc_{x_2^2} C_6^\Delta= \sum_{\{s\} = \pm 1} s_1 s_2 C_6^\Delta(s_1 x_1, s_2 x_2, x_3; \{\vec y\}).
\label{trinaglecut2}
\end{equation}

Combining \eqref{eq:cct} and \eqref{trinaglecut2} then implies that a simple linear combination of the correlator summed over sign flips of certain energy variables should be proportional to the off-shell triangle diagram in flat space, 
which can be expressed in terms of the Bloch-Wigner dilogarithm \cite{Chavez:2012kn}. In fact, a lengthy expression in terms of dilogarithms was recently for found for this correlator \cite{Pimentel:2026kqc}. This implies that the integrals over auxiliary energies in \eqref{eq:cct} must not increase the transcendentality of the flat space result, which is quite non-trivial but in agreement with a conjecture in \cite{Chowdhury:2025ohm}.

Next consider a triple discontinuity involving two $y$ variables and one $x$ variable. While the latter is implemented by cutting one auxiliary propagator, the former correspond to a double discontinuity of the flat space triangle diagram which is computed by cutting all three internal propagators \cite{Britto:2024mna, abreu:2015cuts}, so the triple discontinuity of the correlator is obtained by performing four cuts of the dressed Feynman diagram:  
\begin{align}
&\disc_{x_1^2} \disc_{y_1^2}\disc_{y_2^2} C_6^\Delta = \begin{tikzpicture}[baseline, shift = {(0, 0.1)}, scale=0.5]
\coordinate (A) at (-1, -0.5);
\coordinate (B) at (1, -0.5);
\coordinate (C) at (0, 1);
\coordinate (N) at (1.5, 1.3);
\coordinate (Aleft) at (-2, -0.5);   
\coordinate (Adown) at (-1.5, -1.25); 
\coordinate (Bright) at (2, -0.5);    
\coordinate (Bdown) at (1.5, -1.25);  
\coordinate (Cright) at (0.5, 1.866); 
\coordinate (Cleft) at (-0.5, 1.866); 
\draw (A) -- (B);
\draw (B) -- (C);
\draw (C) -- (A);
\draw (A) -- (Aleft);   
\draw (A) -- (Adown);   
\draw (B) -- (Bright);  
\draw (B) -- (Bdown);   
\draw (C) -- (Cright);  
\draw (C) -- (Cleft);   
\draw[dashed] (N) -- (A);
\draw[dashed] (N) -- (B); 
\draw[dashed] (N) -- (C);
\draw (0.75,0.85) [red]-- (0.7,1.5);
\draw (-0.25,0.08) [red]-- (-0.75,0.42);
\draw (0,-0.25) [red]-- (0,-0.75);
\draw (0.25,0.08) [red]-- (0.75,0.42);
\end{tikzpicture} \\ \nonumber
&=
\Bigg.\sum_{\gamma} \frac{c_\gamma}{x_1 x_2 y_1} \frac{ \tan ^{-1}\left(\frac{-\sqrt{a^2-2 a p+b^2+p^2}+\gamma +p}{\sqrt{-a^2-2 a \gamma -b^2-\gamma ^2}}\right)}{\sqrt{-a^2-2 a \gamma -b^2-\gamma ^2}} \Bigg|_{p = y_2 - x_1}^{p = y_2 + x_1},
\end{align}
where we sum over the following four terms for $\gamma$:
\begin{equation}
c_\gamma \equiv
\begin{cases}
\displaystyle
\frac{1}
{x_1\left(\gamma-x_2+x_3\right)\left(\gamma+x_2+x_3\right)},
& \gamma=\pm x_1, \\[1.2ex]
\displaystyle
\frac{1}
{x_2\left(\gamma+x_1\right)\left(\gamma-x_1\right)},
& \gamma=\pm(x_2-x_3),
\end{cases}
\end{equation}
and 
\begin{eqn}
a = \frac{x_3 y_1 c}{y_2}, \quad 
b^2 = \frac{y_1^2(x_3^2+y_2^2)(1 - c^2)}{y_2^2}, \quad c = \frac{\vec y_1 \cdot \vec y_2}{y_1 y_2}.
\end{eqn}
In this derivation the variable $\vec y_1 \cdot \vec y_2$ is held fixed \cite{Abreu:2014cla}. Finally, let us comment on how the sum rule in \eqref{sumrule} is realised in this example. After integrating out the auxiliary propagator energies and the energy of the loop momentum in \eqref{triangle}, we are left with a three dimensional loop integral whose integrand is given in equation (B.9) of \cite{Chowdhury:2023arc}. Note that this is a very lengthy expression but it is straightforward to check that it vanishes after summing over all sign flips of the $x$-variables according to \eqref{sumrule}. It follows that the same must hold after integration, so it would be interesting to check if previous formulae in the literature \cite{Chowdhury:2023khl, Pimentel:2026kqc} are consistent with this constraint.

\section{Dispersion relations}\label{sec:dispersion}
The dressing rules provide an explicit analytic continuation of each in-in graph to the complex plane of external and internal energies, across regions in which the auxiliary integrals are convergent. For two-site graphs, such as the tree-level diagram in $\phi^3$ theory or the bubble graph in the $\phi^4$ theory, the corresponding analytic region is the entire complex plane of $y^2$, except for possible poles or branch cuts along the positive real axis (assuming $x_1,x_2$ are both real and positive). For $C^\circ_4$ this is manifest in the explicit expression \eqref{eq:lbi}. Leveraging this analyticity property, we can use the Cauchy theorem in the $y^2$ plane to write dispersion relations of the form  
\begin{align}
\nonumber
    F_2(y^2;x_1,x_2)&=\dfrac{1}{2\pi i}\int_0^\infty \dfrac{dz^2}{z^2-y^2}\text{disc}_{z^2}F_2(z^2;x_1,x_2)\\ \label{dispersive}
    & +\text{boundary term}\,,
\end{align}
where $F_2(y^2;x_1,x_2)$ stands for a generic 2-site diagram.  
We have also included a possible contribution from the arc at infinity, $|z^2|\to \infty$, which we refer to as a boundary term. For concreteness, we will assume $\text{Im}(y^2)<0$. While the above integral representation of the correlator $F_2$ requires knowledge of $\text{disc}_{z^2}F_2(z+i\epsilon)$, the latter can be simply read off from the dressing rules, whenever they are available.  More generally, analogous dispersion relations can be derived from the in-in formalism and the requirement of microcausality for certain composite operators which characterise the internal lines. As we will show in Appendix \ref{sec:causality}, taking the discontinuity with respect to $y^2$ is equivalent to replacing all the composite operators' in-in propagators with their non-time-ordered counterparts, effectively cutting all the internal lines in the original graph. 
See also \cite{Tong:2021wai,Ema:2024hkj, Meltzer:2020qbr, Meltzer:2021bmb,Meltzer:2021zin,Salcedo:2022aal, AguiSalcedo:2023nds,Qin:2023nhv, Liu:2024xyi} for related discussions. 

\begin{figure}
\centering
\includegraphics[scale=0.4]{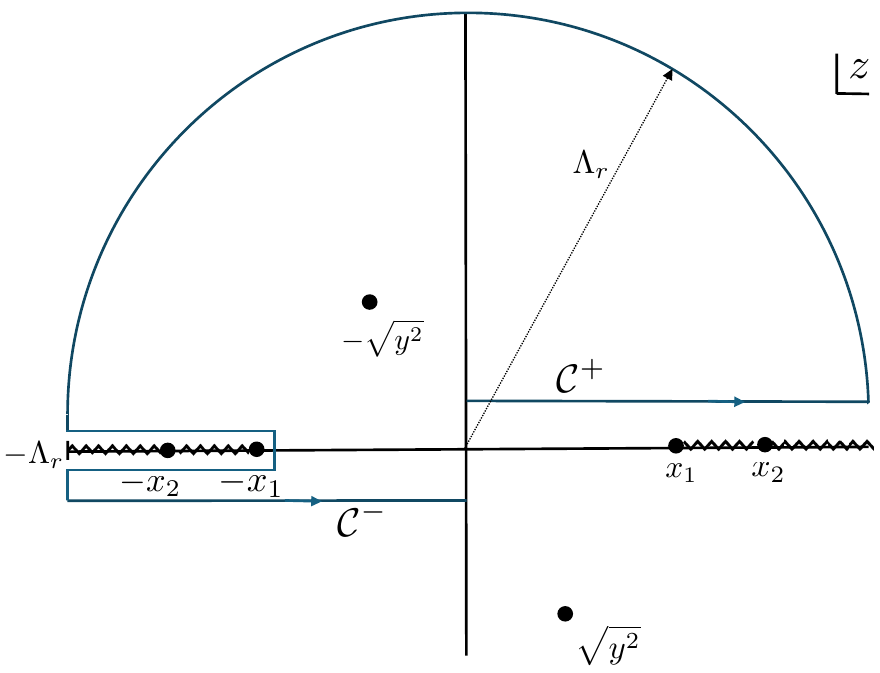}
\caption{The contour integral corresponding to Eq.~\eqref{dispersive2}. Note that, for two-site graphs, the integrand in Eq.~\eqref{dispersive1.5} is analytic in the $z$-plane except for branch cuts and/or poles corresponding to the folded and partial energy singularities, in addition to two simple poles at $z=\pm \sqrt{y^2}$.}
    \label{fig:contour}
\end{figure}

While we used analyticity in the $z^2$ plane to arrive at the dispersive integral in \eqref{dispersive}, the result can be further simplified using the integrand's analyticity in the complex plane of $z=\sqrt{z^2}$. The only possible non-analyticities in the $z-$plane correspond to partial energy singularities of $\text{disc}_{z^2}F_2(z^2;x_{1,2})$ on the real axis, and two simple poles at $z=\pm \sqrt{y^2}$ due to the $z/(z^2-y^2)$ factor (see Fig.~\ref{fig:contour}). For $C^\circ_4$ this is manifest in \eqref{bubmaxdiscn}. 

Employing the symmetry of the integrand   $\frac{2z}{z^2-y^2}\text{disc}_{z^2}F(z^2;\lbrace x\rbrace)$ under $z\to -z$, \eqref{dispersive} can be recast as
\begin{align}
\nno 
    F_2(y^2; x_1,x_2)&=\dfrac{1}{2\pi i}\underset{{\cal C}^-\cup\, {\cal C}^+}{\int}\dfrac{z\,dz}{z^2-y^2}\text{disc}_{z^2}F_2(z^2;x_1,x_2)\\ \label{dispersive1.5}
    & +\text{boundary term}\,,
\end{align}
where ${\cal C}^\pm$ are the semi-infinite contours depicted in Figure~\ref{fig:contour} (see \cite{Meltzer:2021zin,Jazayeri:2021fvk} for similar manipulations of massless wavefunction coefficients). 
Closing the integration contour ${\cal C}^-\cup {\cal C}^+$ in the upper half plane and going around the branch cut along the negative real axis, we finally arrive at 
\begin{align}
\nno
    &F_2(y^2;x_1,x_2)=\\ \nno 
    &-\dfrac{1}{2\pi i}\int_{-\infty}^{-x_1}\dfrac{z\,dz}{z^2-y^2}\text{disc}_z\,\text{disc}_{z^2}F_2(z^2;x_1,x_2)\\ \nno
    &-\sum_{i=1,2}\dfrac{x_i}{x_i^2-y^2}\text{Res}\lbrace\text{disc}_{z^2}F_2(z^2;x_1,x_2)\rbrace\Big|_{z=-x_i} \\ 
    \label{dispersive2}
    &-\dfrac{1}{2}\text{disc}_{y^2}F_2(y^2;x_1,x_2)+\text{boundary term}\,,
\end{align}
which notably includes the sequential discontinuity $\text{disc}_z\,\text{disc}_{z^2}F(z^2;x_1,x_2)$ across the partial energy cuts along $z\in (-\infty,-x_1]$ (assuming $x_1<x_2$ for concreteness). The second term above captures contributions from two possible poles at the partial energy singularities $z=-x_{1,2}$, although they are not present in the examples considered below. Finally, the third term originates from the pole at $z=-y$.

As an illustration, consider the exchange diagram $D_4^{\symbex}$ in \eqref{dressforD2}. Using the dressing rules, the discontinuity across the branch cut $y^2>0$ evaluates to \eqref{bubmaxdisc}, from which we obtain
\begin{align}
&\text{disc}_z\,\text{disc}_{z^2}D_4^{\symbex}(z^2;x_1,x_2)=\\ \nno 
    &\dfrac{-i\pi}{z}\left[\log\left(\dfrac{-z-x_1}{x_1-z}\right)+\log\left(\dfrac{-z-x_2}{x_2-z}\right)\right]\theta(-z-x_2)\\ \nno
    &-\dfrac{i\pi}{z}\log\left(\dfrac{-z-x_2}{x_2-z}\right)\theta(-z-x_1)\theta(z+x_2)\,.
\end{align}
Due to its lower transcendentality, the double discontinuity above is much easier to integrate than the original integrand \eqref{dispersive}. Plugging it back into \eqref{dispersive2}, one can verify that the result agrees with the explicit calculation in \eqref{D2explicit}. Note that the dispersive representation of $D_4^{\symbex}$ in \eqref{dispersive} is actually identical to the original representation in terms of dressing rules in \eqref{dressforD2}. As the next example illustrates, this identification turns out to be only a tree-level accident, with the dispersive and dressing expressions being qualitatively different at loop order. 

Next let us consider the bubble graph in conformally coupled $\phi^4$ theory. The corresponding sequential discontinuity is given by 
\begin{align}
\nno
&\text{disc}_z\,\text{disc}_{z^2}C^{\circ}_4(z^2;x_1,x_2)=-\dfrac{i\pi}{x_1 x_2(x_1+x_2)}\theta(-z-x_2)\\ 
    &+\dfrac{i\pi}{x_1(x_1^2-x_2^2)}\theta(z+x_2)\theta(-z-x_1)\,.
\end{align}
Using this expression, 
an ultraviolet (UV) divergence arises in the first term in \eqref{dispersive2}, which needs to be regularised. To tame the UV behaviour, we deform the infinite contour in ~\eqref{dispersive1.5} to a finite-size contour characterised by the radius $\Lambda_r$, where $\Lambda_r\gg |y|$ is an arbitrary UV scale; see Fig.~\ref{fig:contour}. Performing the integral in \eqref{dispersive1.5} along the new contour then gives   
\begin{align}\label{dispersivebubble}
    &C_4^{\circ}(y^2;x_1,x_2)=\dfrac{1}{2 x_1 x_2(x_1^2-x_2^2)}\times\\ 
    &\left[x_1 \log\left(\dfrac{x_2+y}{\Lambda_r}\right)-x_2 \log\left(\dfrac{x_1+y}{\Lambda_r}\right)\right]+B_{\Lambda_r}(x_1,x_2)\,,\nno
\end{align}
which agrees with the explicit calculation \eqref{bubhardcutoff},  up to the boundary term $B_{\Lambda_r}$. Since this boundary term is independent of $y$, the diagram's $y-$cut is not sufficient for its reconstruction. On the other hand, we may fix $B_{\Lambda_r}$ by demanding that $C_4^{\circ}$ is independent of $\Lambda_r$ and has the correct flat space limit, which is computed from the residue of the correlator at the total-energy singularity $x_1+x_2\to 0$ \cite{Raju:2012zr}. In the end, we find that
\begin{align}
    B_{\Lambda_r}=\dfrac{1}{2x_1 x_2(x_1+x_2)}\log\left(\dfrac{\Lambda_{r}}{2\Lambda}\right)\,.
\end{align}
Combining this with \eqref{dispersivebubble} then reproduces \eqref{bubhardcutoff}. To obtain a dS-invariant expression one needs to work with a regularization scheme that preserves conformal invariance. This can be done by imposing a cutoff on the physical momentum \cite{Senatore:2009cf} or by analytic regularization \cite{Chowdhury:2023arc} and in practice, for this example, it amounts to replacing $\Lambda^2 \to   (x_1 + x_2)^2$ in 
\eqref{bubhardcutoff}. This shall be discussed in more detail in \cite{long}.
\section{Conclusion} \label{conclusion}
Cosmological correlators are objects of great theoretical and observational importance because they provide a direct window into the structure of the early universe. They are traditionally computed using the Scwhinger-Keldysh formalism or by squaring the cosmological wavefunction and performing a path integral over boundary values. On the other hand, in-in correlators in many theories have beautiful mathematical structure which is obscured by these approaches and can be made manifest by applying certain dressing rules to Feynman diagrams in flat space, for which a vast arsenal of computational techniques is available. 

Motivated by this, we have taken the first steps to compute discontinuities of cosmological correlators directly from flat space amplitudes, bypassing many of the complexities that arise in more traditional approaches. The results of this study are very simple to state: discontinuities with respect to internal (or $y$-type) energy variables can be directly uplifted from flat space by dressing unitarity cuts of Feynman diagrams, and discontinuities with respect to external (or $x$-type) energy variables are obtained by cutting auxiliary propagators attached to flat space Feynman diagrams. Cutting all auxiliary propagators in this way leads to a new way to obtain the flat-space limit, including an energy-conserving delta function, via sequential discontinuities. This in turn implies simple sum rules that are obeyed by correlators but not wavefunction coefficients, further demonstrating the advantages of working directly with correlators. Moreover, we demonstrate how to reconstruct correlators from their discontinuities using dispersion relations. This suggests a simple algorithm for computing cosmological correlators: first uplift their discontinuities from flat space, then reconstruct them from dispersion. We illustrate how this works for a number of examples at tree-level and 1-loop in conformally coupled theories for simplicity, although this approach can also be extended to massless theories which exhibit infrared divergences, as we will spell out in a longer publication \cite{long}. Treating IR divergences of cosmological correlators is an important question for which a number of strategies have been proposed \cite{Starobinsky:1986fx, Rajaraman:2010xd, Beneke:2012kn,  Gorbenko:2019rza,Bzowski:2023nef}.

There are several immediate directions for further inquiry, which we hope to address in \cite{long}. In the context of scattering amplitudes, the relation between discontinuities and unitarity cuts follows from the largest time equation \cite{Veltman:1963th, tHooft:1973wag}. Using the dressing rules, we will show that this admits a natural extension to correlators by allowing the cuts to run over both Feynman propagators and auxiliary propagators, unifying the $x$ and $y$ discontinuities into a single framework. The largest time equation was also previously explored in the context of AdS/CFT \cite{Meltzer:2020qbr}. We will also show that the auxiliary propagators appearing in the dressing rules can be put on an equal footing to those of Feynman diagrams by uplifting them to four-dimensional propagators. Applying the largest time equation to this new manifestly four-dimensional representation then implies new cutting rules with respect to Lorentz-invariant variables on which the correlators depend in appropriate kinematics. Combining this with the discontinuities and sum rules derived in this paper then implies powerful constraints which can in principle be used to reconstruct the 1-loop triangle correlator discussed in section \ref{sec:examples} and recently computed in \cite{Pimentel:2026kqc}. It would also be of interest to generalize this to compute 1-loop box correlators. 
Finally, generalising the analysis of two-vertex graphs in Appendix \ref{sec:causality}, we will conduct a more systematic study of the cutting rules implied by microcausality, and their associated dispersion relations in \cite{long}. This complementary viewpoint would enable us in particular to bootstrap more general graphs for which a dressing representation is not known. 

Further down the line, it would be of great interest to extend the differential equations recently developed for wavefunction coefficients \cite{Arkani-Hamed:2023kig,Caloro:2023cep,Grimm:2024tbg} to correlators. Indeed, given their relation to scattering amplitudes it seems very likely that differential equations for flat space Feynman integrals \cite{Henn:2014qga} and other recent technology such as twisted cohomology \cite{Mastrolia:2018uzb,Mizera:2019ose,Caron-Huot:2021xqj} can be uplifted to cosmological correlators. This would in turn suggest the following question: is there an analogue of generalised unitarity for comsological correlators analogous to that which has been so successfully applied to scattering amplitudes \cite{Bern:2011qt}? For example, can one express all 1-loop scalar correlators in a basis of dressed Feynman integrals and determine the coefficients from unitarity cuts? The fact that discontinuities of cosmological correlators can be directly uplifted from those of scattering amplitude in flat space suggests a relation between flat space holography and de Sitter holography. Elucidating such a relation is perhaps the most profound question suggested by this work, and we hope to explore that in the future.  

\begin{acknowledgments}
We thank Joydeep Chakravarty, Ashoke Sen, David Stefanyszyn, Xi Tong, Denis Werth, and Yuhang Zhu for helpful discussions. We thank the authors of \cite{Das:2025qsh} for sharing an upcoming draft with us. CC thanks participants of AdS/CFT Meets Carrollian \& Celestial Holography, Edinburgh; La Ricotta Summer School on The Disordered Universe, Galicia and IISER Bhopal String Group for useful discussions. A.L. is supported by an STFC Consolidated Grant ST/T000708/1. C.C. is supported by an STFC consolidated grant (ST/X000583/1). I.S. is supported by the Excellence Cluster Origins of the DFG under Germany's Excellence Strategy EXC-2094 390783311 as well as EXC 2094/2: ORIGINS 2. S.J. is supported by the STFC Consolidated Grant
ST/X000575/1 and a Simons Investigator Award 690508. J. Marshall is supported by a Durham Doctoral Teaching Fellowship. J. Mei is supported by the European Union (ERC, UNIVERSE PLUS, 101118787). 
\end{acknowledgments}


\appendix

\section{Cutting rules from microcausality} \label{sec:causality}
In this appendix, we use the in-in formalism together with microcausality to derive a set of $y$-cuts, similar to those presented above, although different in the kinematic region across which they apply, as specified below. Crucially, our derivation will not assume any dressing representation for diagrams. Therefore, the final cutting rules are satisfied by more general graphs, with arbitrary external and internal masses and vertex structures.  In particular, see \cite{Tong:2021wai, Jazayeri:2025vlv} for similar considerations at tree-level, and \cite{Ema:2024hkj} for a systematic study of loop-level graphs based on the causal structure of the Schwinger-Keldysh formalism. \\

\noindent \textbf{Derivation:} 
For simplicity, we will concentrate on two-vertex, four-point melon graphs with single-exchange and bubble graphs as special cases. Generalisation to any two-site graph (with an arbitrary number of internal vertices) will be presented elsewhere \cite{long}. We also limit ourselves to vertex structures of the form 
\begin{align}
    {\cal L}_I=-\sqrt{-g}\,\phi^2\,{\cal O}\,,\qquad \text{with}\qquad {\cal O}=\chi^n\,,
\end{align}
where $\phi$ is the external conformally coupled field, and $\chi$ is a field with general mass flowing through internal lines of the melon diagram, as depicted below for $n=3$. 
\begin{align}
\nno
\begin{tikzpicture}[baseline, shift = {(0, 0.25)}]
\node at (-1, -0.25) {$\phi$};
\node at (-0.25, +0.64) {$\chi$};
\draw (-1, 0.5) -- (-0.5, 0);
\draw (-1, -0.5) -- (-0.5, 0);
\draw (1, 0.5) -- (0.5, 0);
\draw (1, -0.5) -- (0.5, 0);
\draw[decorate,decoration={coil,amplitude=1pt, segment length=3pt, aspect=0}] (-0.5,0) -- (0.5,0);
\draw [decorate,decoration={coil,amplitude=1pt, segment length=3pt, aspect=0}] (0, 0) circle (0.5);
\draw[thick] (0,-0.8) [red]-- (0,0.8);
\end{tikzpicture} 
\end{align}

We treat $\chi$ as a distinct field from the external $\phi$ field, but generalisation to the single-field case is straightforward.  The in-in expression for the melon graph can be organized in terms of the two-point function of the \textit{composite operator} ${\cal O}=\chi^n$, using which we may write
\begin{align}
   &\langle \phi(\vec{k}_1)\dots \phi(\vec{k}_4)\rangle=\\ \nno
   &\dfrac{\eta_0^4}{2k_1\dots k_4}F_2(x_1,x_2,y)(2\pi)^3\,\delta^3(\sum \vec{k}_i)\,,
\end{align}
where
\begin{align}
\label{Fcomposite}
   &F_2(x_1,x_2,y)=\sum_{\pm} F_2^{\pm}
   =\sum_{\pm\pm}\dfrac{(\pm i)(\pm i)}{2}\\ \nno
   &\times \int_{-\infty(1\mp i\epsilon)}^{\eta_0} \int_{-\infty(1\mp i\epsilon)}^{\eta_0}\dfrac{d\eta}{\eta^2}\dfrac{d\eta'}{\eta'^2}e^{\pm i x_1\eta}e^{\pm ix_2\eta'}G^{\cal O}_{\pm\pm}(y,\eta,\eta')\,,
\end{align}
$x_1=|\vec{k}_1|+|\vec{k}_2|$, $x_2=|\vec{k}_3|+|\vec{k}_4|$, $y=|\vec{k}_1+\vec{k}_2|$, and $\eta_0$ denotes the end of inflation in conformal time.  The in-in propagators of the composite operator $G^{\cal O}_{\pm\pm}$ are defined as per usual, for instance
\begin{align}
    G_{++}(y,\eta,\eta')=\int d^3\vec{x}'\,e^{-i\vec{y}\cdot\vec{x}'}\,\langle\text{in}|\mathbb{T}\lbrace {\cal O}(\eta,0){\cal O}(\eta',\vec{x}')\rbrace|\text{in}\rangle\,.
\end{align}
As we will show, the compact form of \eqref{Fcomposite}---without expanding the composite operators---will make the analytic properties of $F_2$ due to causality manifest.   

Let us now discuss in more detail the analytic structure of $F_2$ as a function of the complex variable $y^2$, at fixed $x_{1,2}>0$. By employing the K\"{a}ll\'{e}n--Lehmann spectral representation of the two-point $\langle {\cal O}{\cal O}\rangle$, it can be shown that $F_2$ inherits the same analytic structure as a tree-graph with an intermediate fundamental field---we will give a more detailed proof of this in \cite{long}, but see \cite{deRham:2025mjh, Xianyu:2022jwk} for similar discussions. In particular:\\ \\
\textit{$F_2$ can be analytically continued to the complex plane of $y^2$ except for a branch cut along the negative real axis $y^2<0$}. \\ \\
Note that here we are assuming a different analytic continuation than the one in the main text, where the branch cut was placed on the positive real line $y^2>0$. Crucially, in this Appendix, we instead put the branch cut on the negative real axis (e.g., for $\sqrt{y^2}$), in order to avoid partial energy singularities for $F_2$ on the first Riemann sheet. These singularities, which are located at $x_1+\sqrt{y^2}=0$ and $x_2+\sqrt{y^2}=0$, together with their corresponding branch cuts, are hence swept under the second Riemann sheet.    

The next step is to prove that only the Wightman part of the in-in graph, given by Eq.~\eqref{wightmancomponent}, can ever cause discontinuities with respect to $y^2$.
By contrast, the retarded propagator, Eq.~\eqref{retardedfourier}, associated with the composite operator $\chi^n$, would only produce analytic terms in $y^2$ once substituted into \eqref{Fcomposite}. The proof of this step builds on the following three observations:\\ \\
(1) Each two-site diagram can be decomposed into a retarded and a Wightman component. This can be achieved using the following relationships between the (anti-)time-ordered and retarded propagators: 
\begin{align}
    G_{++}(y,\eta,\eta')&=G_R(y,\eta,\eta')+G_{+-}(y,\eta,\eta')\,,\\ \nno
    G_{--}(y,\eta,\eta')&=-G_R(y,\eta,\eta')+G_{-+}(y,\eta,\eta')\,,
\end{align}
where 
\begin{align}
    \nno 
    &G^{{\cal O}}_R(y,\eta,\eta')\\ \label{retardedfourier}
    &=\int d^3\vec{x}'\,e^{-i\vec{y}\cdot \vec{x}'}\lin [{\cal O}(\eta,0),{\cal O}(\eta',\vec{x}')]\rin\theta(\eta-\eta')\,.
\end{align}
Plugging the above decompositions into ~\eqref{Fcomposite} then gives
\begin{align}
\nno 
    F_2=F^R_2+F^W_2\,,
\end{align}
where 
\begin{align}
\label{Fretarded}
    &F_2^R(x_1,x_2,y^2)=\\ \nno 
    &-\dfrac{1}{2}\int\dfrac{d\eta}{\eta^2}\dfrac{d\eta'}{\eta'^2}e^{+i x_1\eta}e^{+ ix_2\eta'}G^{\cal O}_{R}(y,\eta,\eta')\\ \nno
    &+\dfrac{1}{2}\int \dfrac{d\eta}{\eta^2}\dfrac{d\eta'}{\eta'^2}e^{-i x_1\eta}e^{- ix_2\eta'}G^{\cal O}_{R}(y,\eta,\eta')\,,
\end{align}
and 
\begin{align}
\label{wightmancomponent}
    F_2^W&=F_2^{+-}+F_2^{-+}\\ \nno 
&-\dfrac{1}{2}\int^{\eta_0}_{-\infty} \dfrac{d\eta}{\eta^2}\dfrac{d\eta'}{\eta'^2}e^{i x_1\eta}e^{ix_2\eta'}G^{\cal O}_{+-}(y,\eta,\eta')\\ \nno
&-\dfrac{1}{2}\int_{-\infty}^{\eta_0} \dfrac{d\eta}{\eta^2}\dfrac{d\eta'}{\eta'^2}e^{-i x_1\eta}e^{- ix_2\eta'}G^{\cal O}_{-+}(y,\eta,\eta')\,.
\end{align}
Most notably, each term in the last expression consists only of non-time-ordered propagators and therefore factorises into a product of single time integrals. \\ \\
(2) At fixed conformal times $\eta$ and $\eta'$, the Fourier transformation of the retarded propagator $G^{{\cal O}}_R(y,\eta,\eta')$ according to the Paley-Wiener theorem (e.g. \cite{Hui:2025aja, Ema:2024hkj}) is an \textit{entire function} of $y^2$. This is a direct consequence of microcausality, which enforces the commutator $[{\cal O}(\eta,0),{\cal O}(\eta',\vec{x}')]$ to vanish outside the light-cone, $(\eta-\eta')^2<|\vec{x}'|^2$, thereby rendering the integral \eqref{retardedfourier} convergent for an arbitrary complex momentum $\vec{y}$.\\ \\
(3) The retarded component of the graph, defined in \eqref{Fretarded}
is analytic over the complex plane of $y^2$ except for poles and/or branch cuts along $[\text{min}(x_1,x_2),+\infty)$. This property follows from (1), alongside the absence of partial energy singularities on the first sheet. More concisely, we have 
\begin{align}
    \text{disc}_{y^2}F_2^R=0\,\quad \text{if}\quad y^2\in \mathbb{R}^-\,, x_{1,2}\in \mathbb{R}^+\,.
\end{align}
From the properties (1),(2) and (3), we conclude the following cutting rule:  
\begin{align}
\label{causalitycut}
\text{disc}_{y^2}F_2=\text{disc}_{y^2}F^W_2\, \quad \text{if}\quad y^2\in \mathbb{R}^-\,, x_{1,2}\in \mathbb{R}^+\,.
\end{align}
Therefore, the implication of micro-causality for two-site graphs can be summarised as follows: \\ \\
\textit{Knowing the Wightman component of a two-site graph, i.e. \eqref{wightmancomponent}, is sufficient to reconstruct its discontinuity across $y^2<0$.} \\

Once the discontinuity across the branch cut along $y^2<0$ is computed using \eqref{causalitycut}, one can build similar dispersion relations to the one discussed in section \ref{sec:dispersion} in order to obtain the full diagram \cite{long}. \\ 

\noindent \textbf{The bubble graph example:} To illustrate the causality-based cutting rule \eqref{causalitycut}, we consider the bubble graph in the $\phi^2\chi^2$ theory, taking $\chi$ to be conformally coupled. 
Despite differing by a $ y$-independent prefactor, we choose the same notation $C_4^\circ$ as the $\phi^4$ bubble graph in the preceding sections to denote the $\phi^2\chi^2$ bubble diagram.  

Since the Wightman piece involves a loop integral, the computation will be more involved. So let us break it down by focusing on the second line in \eqref{wightmancomponent}, hereafter called $C_4^{++W}$. Expanding the composite $G_{+-}$ propagator via Wick's theorem, we find \footnote{In this equation, a technical subtlety arises in changing the order of the integrals over time and the loop momentum. For instance, in \eqref{ppW}, the second time integral diverges for $x_2-|\vec{l}|-|\vec{l}+\vec{y}|<0$. To remedy this problem, we first consider loop momenta for which this time integral is convergent and, by setting $x_2\to x_2-i\epsilon$, analytically continue the result to arbitrary $\vec{l}$.}:  
\begin{align}
\label{ppW}
    &C_4^{++W}=-\dfrac{1}{2}\int \dfrac{d^3\vec{l}}{(2\pi)^3} \\ \nno
    &\left[\int \dfrac{d\eta}{\eta^2}\,e^{ix_1\eta}\chi_+(|\vec{l}|,\eta)\chi_+(|\vec{y}+\vec{l}|,\eta)\right]\\ \nno
    &\times \left[\int \dfrac{d\eta'}{\eta'^2}\,e^{ix_2\eta'}\chi_-(|\vec{l}|,\eta')\chi_-(|\vec{y}+\vec{l}|,\eta')\right]\,\\ \nno 
    &=-\dfrac{1}{8}\int \dfrac{d^3 l}{|\vec{l}|}\dfrac{1}{|\vec{l}+\vec{y}|}\dfrac{1}{x_1+|\vec{l}|+|\vec{l}+\vec{y}|}\dfrac{1}{x_2-|\vec{l}|-|\vec{l}+\vec{y}|-i\epsilon}\,.
\end{align}
The integral over the loop momentum is UV convergent and evaluates to: 
\begin{align}
    C_4^{++W}=-\dfrac{1}{32\pi^2 (x_1+x_2)}\left[i\pi+\log\left(\dfrac{x_2-\sqrt{y^2}}{x_1+\sqrt{y^2}}\right)\right]\,.
\end{align}
The remaining contributions to the Wightman part in \eqref{wightmancomponent} can be similarly computed to yield:  
\begin{align}
    C_4^W &=-\dfrac{1}{32\pi^2}\left[\dfrac{1}{x_1-x_2}\log\left(\dfrac{x_1+\sqrt{y^2}}{x_2+\sqrt{y^2}}\right)\right.\\ \nno 
    &\qquad \qquad \quad+\left.\dfrac{1}{x_1+x_2}\log\left(\dfrac{x_2-\sqrt{y^2}}{x_1+\sqrt{y^2}}\right)\right]\,.
\end{align}
As is clear from the above expression, the Wightman component is not symmetric under the exchange of $x_1\leftrightarrow x_2$, unlike the full result $C_4^W+C_4^R$. 

With the graph's Wightman component at hand, using \eqref{causalitycut} we eventually arrive at the total discontinuity with respect to $y^2<0$: 
\begin{align}
\nno 
\text{disc}_{y^2}C^{\circ}_4 =-\dfrac{1}{16\pi^2(x_1^2-x_2^2)}&\left[x_2\log\left(\dfrac{x_1+\sqrt{y^2}}{x_1-\sqrt{y^2}}\right)\right.\\  
    &\left.\qquad -x_1 \log\left(\dfrac{x_2+\sqrt{y^2}}{x_2-\sqrt{y^2}}\right) \right]\,,
\end{align}
which agrees with the $y-$cut of the explicit result in \eqref{bubmaxdiscn}. 

\section{Analytic Regularization}\label{app:dSinv}
The 1-loop bubble example in section \ref{sec:examples} is divergent and requires regularisation. A dS-invariant regulator is implemented by the following modification of \eqref{eq:bub} \cite{Chowdhury:2025ohm}: 
\begin{align}\label{eq:c2_reg}
 & C_4^\circ= \frac{1}{\Gamma(\e)^2} \intsinf ds_1 ds_2 \quad\times \\
&\quad\int\limits_{-\infty}^\infty \frac{dp  \,(s_1s_2)^{\epsilon-1}}{(p^2 - (s_1 + x_1)^2) (p^2 - (s_2 + x_2)^2)}\int \frac{d^{4 - \e} L}{L^2 (L + K)^2} \,.\nno
\end{align}
The $y^2$ discontinuity, which only affects the second integral, is then given by 
\begin{align}
   & \pi^{1-\epsilon}\int\limits _{0}^\infty dl_0 \int\limits _{0}^\infty dl \int\limits _{-1}^1d\cos\theta\frac{\delta(l_0^2-l^2)\delta(K^2-2 l_0\sqrt{K^2})l^{2-2\epsilon}}{\sin(\theta)^{-2\epsilon}\Gamma(1-\epsilon)}\nonumber\\
   &=\theta(p^2-y^2)\pi+O(\epsilon)\,,
\end{align}
which is finite for $\epsilon\to 0$, so that we can set $\epsilon=0$ in the remaining $p$-integral, giving 
\begin{align}
\disc_{y_1^2}C^\circ_4=\frac{\pi}{2}\int\limits_{y}^\infty \frac{dp  \,}{(p^2 -  x_1^2) (p^2 -  x_2^2)}\,,
\end{align}
in agreement with the $y^2$ discontinuity in cut-off regularization \eqref{eq:4pt-ycut}. 

The dS-invariant form of the $x$-cut is interesting since it implies an extra branch cut in the total energy \cite{Raju:2012zr}. To understand this contribution to $C^\circ_4$ we start from the regularized form \eqref{eq:c2_reg}, isolate the simple pole in $\epsilon$ of the last integral, and perform the $p$-integral giving 
\begin{align}
    \frac{1}{\Gamma(\epsilon)}\int\limits_0^\infty& \frac{ ds_1ds_2 \; \,(s_1s_2)^{\epsilon-1}}{s_1+s_2+E}\,,\nno
\end{align}
where $E=x_1+x_2$ is the total energy. Using \eqref{discid} we see that there is a discontinuity for $E<0$ with
\begin{align}
 \disc_{E}C_4^\circ =&   \frac{2i}{\Gamma(\epsilon)}\int\limits_0^\infty  \frac{ds_1ds_2 \;}{(s_1s_2)^{1-\epsilon}}\delta(s_1\!+\!s_2\!+\!E)\sim\frac{1}{E}\,.\nno
\end{align}
This is in agreement with the explicit $\log(E)$-dependence of the explicit  result in \cite{Chowdhury:2023arc}.

\section{Lorentzian bubble Integral}\label{sec:Lbub}
In this Appendix, we derive \eqref{bubmaxdiscn} by evaluating the integral in \eqref{bubmaxdiscn} in complex kinematics and then computing the $y^2$ discontinuity. We assume that $x_1$ and $x_2$ are positive and  $y^2=s+i d $ with $s>0$, $d>0$. We also assume a momentum cut-off. Then
\begin{widetext}
\begin{align}\label{eq:lbi}
C_4^{\circ}=&-\frac{i\pi x_2}{x_1^2 - x_2^2}
\log\!\left(1-\frac{s- id }{ x_1}\right)
+ \frac{i\pi x_1}{x_1^2 - x_2^2}
\log\!\left(1-\frac{s-i d}{x_2}\right) - \frac{i\pi}{x_1 + x_2}\log\left(\frac{-s + i d}{s - i d}\right) 
\\[6pt]
&+ \frac{\sqrt[4]{-1}\,\pi  \sqrt{-s + i d}}{\sqrt{d + i s}\,(x_1^2 - x_2^2)}
\left[x_2
\log\!\left(\frac{1 -\sqrt{d + i s}}{1 +\sqrt{d + i s}}\right)
-x_1 \log\!\left(\frac{1 -\sqrt{d + i s}}{1 +\sqrt{d + i s}}\right)
\right] 
- \frac{i\pi x_2 \log(x_1^2)}{x_1^2 - x_2^2}
+ \frac{i\pi x_1 \log(x_2^2)}{x_1^2 - x_2^2}
- \frac{\pi^2}{x_1 + x_2}\nonumber\,.
\end{align}
\end{widetext}
There is a similar equation for $d<0$. Taking the difference, we reproduce the predicted discontinuity from the cut diagram in equation \eqref{bubmaxdiscn}. Note that there is no contribution from the square roots in the logs since $ y^2$ enters through $i y^2$ there. Instead, the discontinuity come from the prefactors. 

\bibliography{references}

@article{Guth:1980zm,
    author = "Guth, Alan H.",
    editor = "Fang, Li-Zhi and Ruffini, R.",
    title = "{The Inflationary Universe: A Possible Solution to the Horizon and Flatness Problems}",
    reportNumber = "SLAC-PUB-2576",
    doi = "10.1103/PhysRevD.23.347",
    journal = "Phys. Rev. D",
    volume = "23",
    pages = "347--356",
    year = "1981"
}

@article{Starobinsky:1980te,
    author = "Starobinsky, Alexei A.",
    editor = "Khalatnikov, I. M. and Mineev, V. P.",
    title = "{A New Type of Isotropic Cosmological Models Without Singularity}",
    doi = "10.1016/0370-2693(80)90670-X",
    journal = "Phys. Lett. B",
    volume = "91",
    pages = "99--102",
    year = "1980"
}

@article{Mukhanov:1981xt,
      author         = "Mukhanov, Viatcheslav F. and Chibisov, G. V.",
      title          = "{Quantum Fluctuations and a Nonsingular Universe}",
      journal        = "JETP Lett.",
      volume         = "33",
      year           = "1981",
      pages          = "532-535",
      note           = "[Pisma Zh. Eksp. Teor. Fiz.33,549(1981)]",
      SLACcitation   = "%%CITATION = JTPLA,33,532;%%"
}

@article{Linde:1981mu,
    author = "Linde, Andrei D.",
    editor = "Fang, Li-Zhi and Ruffini, R.",
    title = "{A New Inflationary Universe Scenario: A Possible Solution of the Horizon, Flatness, Homogeneity, Isotropy and Primordial Monopole Problems}",
    reportNumber = "LEBEDEV-81-229",
    doi = "10.1016/0370-2693(82)91219-9",
    journal = "Phys. Lett. B",
    volume = "108",
    pages = "389--393",
    year = "1982"
}

@article{Jazayeri:2025vlv,
    author = "Jazayeri, Sadra and Tong, Xi and Zhu, Yuhang",
    title = "{Every Wrinkle Carries A Memory: An Integro-differential Bootstrap for Features in Cosmological Correlators}",
    eprint = "2511.00152",
    archivePrefix = "arXiv",
    primaryClass = "hep-th",
    month = "10",
    year = "2025"
}

@article{Xianyu:2022jwk,
    author = "Xianyu, Zhong-Zhi and Zhang, Hongyu",
    title = "{Bootstrapping one-loop inflation correlators with the spectral decomposition}",
    eprint = "2211.03810",
    archivePrefix = "arXiv",
    primaryClass = "hep-th",
    doi = "10.1007/JHEP04(2023)103",
    journal = "JHEP",
    volume = "04",
    pages = "103",
    year = "2023"
}

@article{Qin:2023nhv,
    author = "Qin, Zhehan and Xianyu, Zhong-Zhi",
    title = "{Nonanalyticity and on-shell factorization of inflation correlators at all loop orders}",
    eprint = "2308.14802",
    archivePrefix = "arXiv",
    primaryClass = "hep-th",
    doi = "10.1007/JHEP01(2024)168",
    journal = "JHEP",
    volume = "01",
    pages = "168",
    year = "2024"
}

@article{Liu:2024xyi,
    author = "Liu, Haoyuan and Qin, Zhehan and Xianyu, Zhong-Zhi",
    title = "{Dispersive bootstrap of massive inflation correlators}",
    eprint = "2407.12299",
    archivePrefix = "arXiv",
    primaryClass = "hep-th",
    reportNumber = "USTC-ICTS/PCFT-24-23",
    doi = "10.1007/JHEP02(2025)101",
    journal = "JHEP",
    volume = "02",
    pages = "101",
    year = "2025"
}

@article{deRham:2025mjh,
    author = "de Rham, Claudia and Jazayeri, Sadra and Tolley, Andrew J.",
    title = "{Bispectrum islands: Bootstrap bounds on cosmological correlators}",
    eprint = "2506.19198",
    archivePrefix = "arXiv",
    primaryClass = "hep-th",
    reportNumber = "Imperial-TP-2025-cdr2, Imperial/TP/2025/cdr2",
    doi = "10.1103/q6rq-sj9t",
    journal = "Phys. Rev. D",
    volume = "112",
    number = "8",
    pages = "083531",
    year = "2025"
}

@article{AguiSalcedo:2023nds,
    author = "Agui Salcedo, Santiago and Melville, Scott",
    title = "{The cosmological tree theorem}",
    eprint = "2308.00680",
    archivePrefix = "arXiv",
    primaryClass = "hep-th",
    doi = "10.1007/JHEP12(2023)076",
    journal = "JHEP",
    volume = "12",
    pages = "076",
    year = "2023"
}

@article{Hui:2025aja,
    author = "Hui, Lam and Nicolis, Alberto and Podo, Alessandro and Zhou, Shengjia",
    title = "{Microcausality without Lorentz invariance}",
    eprint = "2502.04215",
    archivePrefix = "arXiv",
    primaryClass = "hep-th",
    doi = "10.1007/JHEP07(2025)188",
    journal = "JHEP",
    volume = "07",
    pages = "188",
    year = "2025"
}

@article{Glew:2025otn,
    author = "Glew, Ross and Lukowski, Tomasz",
    title = "{Amplitubes: graph cosmohedra}",
    eprint = "2502.17564",
    archivePrefix = "arXiv",
    primaryClass = "hep-th",
    doi = "10.1007/JHEP09(2025)074",
    journal = "JHEP",
    volume = "09",
    pages = "074",
    year = "2025"
}

@article{Glew:2025mry,
    author = "Glew, Ross",
    title = "{Correlators from graphical amplitudes}",
    doi = "10.1103/cg2r-gtkt",
    journal = "Phys. Rev. D",
    volume = "112",
    number = "6",
    pages = "L061302",
    year = "2025"
}

@article{Starobinsky:1986fx,
    author = "Starobinsky, Alexei A.",
    title = "{STOCHASTIC DE SITTER (INFLATIONARY) STAGE IN THE EARLY UNIVERSE}",
    doi = "10.1007/3-540-16452-9_6",
    journal = "Lect. Notes Phys.",
    volume = "246",
    pages = "107--126",
    year = "1986"
}

@article{Pimentel:2026kqc,
    author = "Pimentel, Guilherme L. and Westerdijk, Tom",
    title = "{On Cosmological Correlators at One Loop}",
    eprint = "2601.00952",
    archivePrefix = "arXiv",
    primaryClass = "hep-th",
    month = "1",
    year = "2026"
}

@article{Caloro:2023cep,
    author = "Caloro, Francesca and McFadden, Paul",
    title = "{$\mathcal{A}$-hypergeometric functions and creation operators for Feynman and Witten diagrams}",
    eprint = "2309.15895",
    archivePrefix = "arXiv",
    primaryClass = "hep-th",
    month = "9",
    year = "2023"
}

@article{Grimm:2024tbg,
    author = "Grimm, Thomas W. and Hoefnagels, Arno",
    title = "{Reductions of GKZ systems and applications to cosmological correlators}",
    eprint = "2409.13815",
    archivePrefix = "arXiv",
    primaryClass = "hep-th",
    doi = "10.1007/JHEP04(2025)196",
    journal = "JHEP",
    volume = "04",
    pages = "196",
    year = "2025"
}

@article{Caron-Huot:2021xqj,
    author = "Caron-Huot, Simon and Pokraka, Andrzej",
    title = "{Duals of Feynman integrals. Part I. Differential equations}",
    eprint = "2104.06898",
    archivePrefix = "arXiv",
    primaryClass = "hep-th",
    doi = "10.1007/JHEP12(2021)045",
    journal = "JHEP",
    volume = "12",
    pages = "045",
    year = "2021"
}

@article{Henn:2014qga,
    author = "Henn, Johannes M.",
    title = "{Lectures on differential equations for Feynman integrals}",
    eprint = "1412.2296",
    archivePrefix = "arXiv",
    primaryClass = "hep-ph",
    doi = "10.1088/1751-8113/48/15/153001",
    journal = "J. Phys. A",
    volume = "48",
    pages = "153001",
    year = "2015"
}

@article{Bern:2011qt,
    author = "Bern, Zvi and Huang, Yu-tin",
    title = "{Basics of Generalized Unitarity}",
    eprint = "1103.1869",
    archivePrefix = "arXiv",
    primaryClass = "hep-th",
    reportNumber = "UCLA-11-TEP-103",
    doi = "10.1088/1751-8113/44/45/454003",
    journal = "J. Phys. A",
    volume = "44",
    pages = "454003",
    year = "2011"
}

@inproceedings{Britto:2024mna,
    author = "Britto, Ruth and Duhr, Claude and Hannesdottir, Holmfridur S. and Mizera, Sebastian",
    title = "{Cutting-Edge Tools for Cutting Edges}",
    eprint = "2402.19415",
    archivePrefix = "arXiv",
    primaryClass = "hep-th",
    reportNumber = "BONN-TH-2024-05",
    doi = "10.1016/B978-0-323-95703-8.00097-5",
    month = "2",
    year = "2024"
}

@article{Mastrolia:2018uzb,
    author = "Mastrolia, Pierpaolo and Mizera, Sebastian",
    title = "{Feynman Integrals and Intersection Theory}",
    eprint = "1810.03818",
    archivePrefix = "arXiv",
    primaryClass = "hep-th",
    doi = "10.1007/JHEP02(2019)139",
    journal = "JHEP",
    volume = "02",
    pages = "139",
    year = "2019"
}

@article{Mizera:2019ose,
    author = "Mizera, Sebastian",
    title = "{Status of Intersection Theory and Feynman Integrals}",
    eprint = "2002.10476",
    archivePrefix = "arXiv",
    primaryClass = "hep-th",
    doi = "10.22323/1.383.0016",
    journal = "PoS",
    volume = "MA2019",
    pages = "016",
    year = "2019"
}

@article{Achucarro:2022qrl,
    author = "Ach{\'u}carro, Ana and others",
    title = "{Inflation: Theory and Observations}",
    eprint = "2203.08128",
    archivePrefix = "arXiv",
    primaryClass = "astro-ph.CO",
    month = "3",
    year = "2022"
}

@article{Arkani-Hamed:2025mce,
    author = "Arkani-Hamed, Nima and Glew, Ross and Vaz{\~a}o, Francisco",
    title = "{Correlators are simpler than wavefunctions}",
    eprint = "2512.23795",
    archivePrefix = "arXiv",
    primaryClass = "hep-th",
    month= "12",
    year = "2025"
}

@article{long,
    author = "Chowdhury, Chandramouli and Jazayeri, Sadra and Lipstein, Arthur and Marshall, Joe and Mei, Jiajie and Sachs, Ivo",
    eprint = "{in preparation}"
}

@article{Salcedo:2022aal,
    author = "Salcedo, Santiago Agui and Lee, Mang Hei Gordon and Melville, Scott and Pajer, Enrico",
    title = "{The Analytic Wavefunction}",
    eprint = "2212.08009",
    archivePrefix = "arXiv",
    primaryClass = "hep-th",
    doi = "10.1007/JHEP06(2023)020",
    journal = "JHEP",
    volume = "06",
    pages = "020",
    year = "2023"
}

@article{Chowdhury:2025nnk,
    author = "Chowdhury, Chandramouli and Lipstein, Arthur and Marshall, Joe and Zhang, Alex Jiayi",
    title = "{On in-in correlators for spinning theories and their shadow formulation}",
    eprint = "2512.14694",
    archivePrefix = "arXiv",
    primaryClass = "hep-th",
    month = "12",
    year = "2025"
}

@article{Armstrong:2020woi,
    author = "Armstrong, Connor and Lipstein, Arthur E. and Mei, Jiajie",
    title = "{Color/kinematics duality in AdS$_{4}$}",
    eprint = "2012.02059",
    archivePrefix = "arXiv",
    primaryClass = "hep-th",
    doi = "10.1007/JHEP02(2021)194",
    journal = "JHEP",
    volume = "02",
    pages = "194",
    year = "2021"
}

@article{Albayrak:2020fyp,
    author = "Albayrak, Soner and Kharel, Savan and Meltzer, David",
    title = "{On duality of color and kinematics in (A)dS momentum space}",
    eprint = "2012.10460",
    archivePrefix = "arXiv",
    primaryClass = "hep-th",
    doi = "10.1007/JHEP03(2021)249",
    journal = "JHEP",
    volume = "03",
    pages = "249",
    year = "2021"
}

@article{Chowdhury:2023arc,
    author = "Chowdhury, Chandramouli and Lipstein, Arthur and Mei, Jiajie and Sachs, Ivo and Vanhove, Pierre",
    title = "{The subtle simplicity of cosmological correlators}",
    eprint = "2312.13803",
    archivePrefix = "arXiv",
    primaryClass = "hep-th",
    reportNumber = "LMU-ASC 37/23, IPhT-t23/119",
    doi = "10.1007/JHEP03(2025)007",
    journal = "JHEP",
    volume = "03",
    pages = "007",
    year = "2025"
}

@article{Chowdhury:2025ohm,
    author = "Chowdhury, Chandramouli and Lipstein, Arthur and Marshall, Joe and Mei, Jiajie and Sachs, Ivo",
    title = "{Cosmological Dressing Rules}",
    eprint = "2503.10598",
    archivePrefix = "arXiv",
    primaryClass = "hep-th",
    reportNumber = "LMU-ASC 03/25",
    month = "3",
    year = "2025"
}

@article{Sleight:2020obc,
    author = "Sleight, Charlotte and Taronna, Massimo",
    title = "{From AdS to dS exchanges: Spectral representation, Mellin amplitudes, and crossing}",
    eprint = "2007.09993",
    archivePrefix = "arXiv",
    primaryClass = "hep-th",
    doi = "10.1103/PhysRevD.104.L081902",
    journal = "Phys. Rev. D",
    volume = "104",
    number = "8",
    pages = "L081902",
    year = "2021"
}

@article{Schaub:2023scu,
    author = "Schaub, Vladimir",
    title = "{Spinors in (Anti-)de Sitter Space}",
    eprint = "2302.08535",
    archivePrefix = "arXiv",
    primaryClass = "hep-th",
    doi = "10.1007/JHEP09(2023)142",
    journal = "JHEP",
    volume = "09",
    pages = "142",
    year = "2023"
}

@article{Sleight:2025dmt,
    author = "Sleight, Charlotte and Taronna, Massimo",
    title = "{(Non-)Conserved Currents and Cosmological Correlators}",
    eprint = "2509.18888",
    archivePrefix = "arXiv",
    primaryClass = "hep-th",
    month = "9",
    year = "2025"
}

@article{Arkani-Hamed:2023kig,
    author = "Arkani-Hamed, Nima and Baumann, Daniel and Hillman, Aaron and Joyce, Austin and Lee, Hayden and Pimentel, Guilherme L.",
    title = "{Differential equations for cosmological correlators}",
    eprint = "2312.05303",
    archivePrefix = "arXiv",
    primaryClass = "hep-th",
    doi = "10.1007/JHEP09(2025)009",
    journal = "JHEP",
    volume = "09",
    pages = "009",
    year = "2025"
}

@article{Arkani-Hamed:2023bsv,
    author = "Arkani-Hamed, Nima and Baumann, Daniel and Hillman, Aaron and Joyce, Austin and Lee, Hayden and Pimentel, Guilherme L.",
    title = "{Kinematic Flow and the Emergence of Time}",
    eprint = "2312.05300",
    archivePrefix = "arXiv",
    primaryClass = "hep-th",
    doi = "10.1103/dsjm-tckw",
    journal = "Phys. Rev. Lett.",
    volume = "135",
    number = "3",
    pages = "031602",
    year = "2025"
}

@article{Melville:2021lst,
    author = "Melville, Scott and Pajer, Enrico",
    title = "{Cosmological Cutting Rules}",
    eprint = "2103.09832",
    archivePrefix = "arXiv",
    primaryClass = "hep-th",
    doi = "10.1007/JHEP05(2021)249",
    journal = "JHEP",
    volume = "05",
    pages = "249",
    year = "2021"
}

@article{Arkani-Hamed:2024jbp,
    author = "Arkani-Hamed, Nima and Figueiredo, Carolina and Vaz{\~a}o, Francisco",
    title = "{Cosmohedra}",
    eprint = "2412.19881",
    archivePrefix = "arXiv",
    primaryClass = "hep-th",
    doi = "10.1007/JHEP11(2025)029",
    journal = "JHEP",
    volume = "11",
    pages = "029",
    year = "2025"
}

@article{Gomez:2021qfd,
    author = "Gomez, Humberto and Jusinskas, Renann Lipinski and Lipstein, Arthur",
    title = "{Cosmological Scattering Equations}",
    eprint = "2106.11903",
    archivePrefix = "arXiv",
    primaryClass = "hep-th",
    doi = "10.1103/PhysRevLett.127.251604",
    journal = "Phys. Rev. Lett.",
    volume = "127",
    number = "25",
    pages = "251604",
    year = "2021"
}

@article{Sleight:2019mgd,
    author = "Sleight, Charlotte",
    title = "{A Mellin Space Approach to Cosmological Correlators}",
    eprint = "1906.12302",
    archivePrefix = "arXiv",
    primaryClass = "hep-th",
    doi = "10.1007/JHEP01(2020)090",
    journal = "JHEP",
    volume = "01",
    pages = "090",
    year = "2020"
}

@article{Jazayeri:2021fvk,
    author = "Jazayeri, Sadra and Pajer, Enrico and Stefanyszyn, David",
    title = "{From locality and unitarity to cosmological correlators}",
    eprint = "2103.08649",
    archivePrefix = "arXiv",
    primaryClass = "hep-th",
    doi = "10.1007/JHEP10(2021)065",
    journal = "JHEP",
    volume = "10",
    pages = "065",
    year = "2021"
}

@article{Bzowski:2019kwd,
    author = "Bzowski, Adam and McFadden, Paul and Skenderis, Kostas",
    title = "{Conformal $n$-point functions in momentum space}",
    eprint = "1910.10162",
    archivePrefix = "arXiv",
    primaryClass = "hep-th",
    doi = "10.1103/PhysRevLett.124.131602",
    journal = "Phys. Rev. Lett.",
    volume = "124",
    number = "13",
    pages = "131602",
    year = "2020"
}

@article{Raju:2011mp,
	archiveprefix = {arXiv},
	author = {Raju, Suvrat},
	doi = {10.1103/PhysRevD.83.126002},
	eprint = {1102.4724},
	journal = {Phys. Rev. D},
	pages = {126002},
	primaryclass = {hep-th},
	reportnumber = {HRI-ST-1103},
	title = {{Recursion Relations for AdS/CFT Correlators}},
	volume = {83},
	year = {2011}}

@article{Bzowski:2013sza,
	archiveprefix = {arXiv},
	author = {Bzowski, Adam and McFadden, Paul and Skenderis, Kostas},
	doi = {10.1007/JHEP03(2014)111},
	eprint = {1304.7760},
	journal = {JHEP},
	pages = {111},
	primaryclass = {hep-th},
	title = {{Implications of conformal invariance in momentum space}},
	volume = {03},
	year = {2014}}

@article{Arkani-Hamed:2017fdk,
	archiveprefix = {arXiv},
	author = {Arkani-Hamed, Nima and Benincasa, Paolo and Postnikov, Alexander},
	eprint = {1709.02813},
	month = "9",
	primaryclass = {hep-th},
	title = {{Cosmological Polytopes and the Wavefunction of the Universe}},
	year = {2017}}

@article{Weinberg:2005vy,
	archiveprefix = {arXiv},
	author = {Weinberg, Steven},
	doi = {10.1103/PhysRevD.72.043514},
	eprint = {hep-th/0506236},
	journal = {Phys. Rev. D},
	pages = {043514},
	reportnumber = {UTTG-01-05},
	title = {{Quantum contributions to cosmological correlations}},
	volume = {72},
	year = {2005}}

@article{Maldacena:2002vr,
	archiveprefix = {arXiv},
	author = {Maldacena, Juan Martin},
	doi = {10.1088/1126-6708/2003/05/013},
	eprint = {astro-ph/0210603},
	journal = {JHEP},
	pages = {013},
	title = {{Non-Gaussian features of primordial fluctuations in single field inflationary models}},
	volume = {05},
	year = {2003}}

@article{Hartle:1983ai,
	author = {Hartle, J. B. and Hawking, S. W.},
	doi = {10.1103/PhysRevD.28.2960},
	editor = {Fang, Li-Zhi and Ruffini, R.},
	journal = {Phys. Rev. D},
	pages = {2960--2975},
	reportnumber = {PRINT-83-0937 (CAMBRIDGE)},
	title = {{Wave Function of the Universe}},
	volume = {28},
	year = {1983}}

@article{Raju:2010by,
	archiveprefix = {arXiv},
	author = {Raju, Suvrat},
	doi = {10.1103/PhysRevLett.106.091601},
	eprint = {1011.0780},
	journal = {Phys. Rev. Lett.},
	pages = {091601},
	primaryclass = {hep-th},
	reportnumber = {HRI-ST-1009},
	title = {{BCFW for Witten Diagrams}},
	volume = {106},
	year = {2011}}

@article{Arkani-Hamed:2015bza,
	archiveprefix = {arXiv},
	author = {Arkani-Hamed, Nima and Maldacena, Juan},
	eprint = {1503.08043},
	month = "3",
	primaryclass = {hep-th},
	title = {{Cosmological Collider Physics}},
	year = {2015}}

@article{Arkani-Hamed:2018kmz,
	archiveprefix = {arXiv},
	author = {Arkani-Hamed, Nima and Baumann, Daniel and Lee, Hayden and Pimentel, Guilherme L.},
	doi = {10.1007/JHEP04(2020)105},
	eprint = {1811.00024},
	journal = {JHEP},
	pages = {105},
	primaryclass = {hep-th},
	title = {{The Cosmological Bootstrap: Inflationary Correlators from Symmetries and Singularities}},
	volume = {04},
	year = {2020}}

@article{Raju:2012zr,
	archiveprefix = {arXiv},
	author = {Raju, Suvrat},
	doi = {10.1103/PhysRevD.85.126009},
	eprint = {1201.6449},
	journal = {Phys. Rev. D},
	pages = {126009},
	primaryclass = {hep-th},
	reportnumber = {HRI-ST-1201},
	title = {{New Recursion Relations and a Flat Space Limit for AdS/CFT Correlators}},
	volume = {85},
	year = {2012}}

@article{Gorbenko:2019rza,
	archiveprefix = {arXiv},
	author = {Gorbenko, Victor and Senatore, Leonardo},
	eprint = {1911.00022},
	month = "10",
	primaryclass = {hep-th},
	title = {{$\lambda \phi^4$ in dS}},
	year = {2019}}

@article{Chowdhury:2023khl,
	archiveprefix = {arXiv},
	author = {Chowdhury, Chandramouli and Singh, Kajal},
	eprint = {2305.18529},
	month = "5",
	primaryclass = {hep-th},
	title = {{Analytic Results for Loop-Level Momentum Space Witten Diagrams}},
	year = {2023}}

@book{Peskin:1995ev,
    author = "Peskin, Michael E. and Schroeder, Daniel V.",
    title = "{An Introduction to quantum field theory}",
    isbn = "978-0-201-50397-5",
    publisher = "Addison-Wesley",
    address = "Reading, USA",
    year = "1995"
}

@article{Benincasa:2024leu,
    author = "Benincasa, Paolo and Dian, Gabriele",
    title = "{The Geometry of Cosmological Correlators}",
    eprint = "2401.05207",
    archivePrefix = "arXiv",
    primaryClass = "hep-th",
    reportNumber = "MPP-2023-150, DESY-24-006",
    month = "1",
    year = "2024"
}

@article{DiPietro:2021sjt,
    author = "Di Pietro, Lorenzo and Gorbenko, Victor and Komatsu, Shota",
    title = "{Analyticity and unitarity for cosmological correlators}",
    eprint = "2108.01695",
    archivePrefix = "arXiv",
    primaryClass = "hep-th",
    reportNumber = "CERN-TH-2021-118",
    doi = "10.1007/JHEP03(2022)023",
    journal = "JHEP",
    volume = "03",
    pages = "023",
    year = "2022"
}

@article{Bzowski:2023nef,
    author = "Bzowski, Adam and McFadden, Paul and Skenderis, Kostas",
    title = "{Renormalisation of IR divergences and holography in de Sitter}",
    eprint = "2312.17316",
    archivePrefix = "arXiv",
    primaryClass = "hep-th",
    month = "12",
    year = "2023"
}

@article{Goodhew:2020hob,
    author = "Goodhew, Harry and Jazayeri, Sadra and Pajer, Enrico",
    title = "{The Cosmological Optical Theorem}",
    eprint = "2009.02898",
    archivePrefix = "arXiv",
    primaryClass = "hep-th",
    doi = "10.1088/1475-7516/2021/04/021",
    journal = "JCAP",
    volume = "04",
    pages = "021",
    year = "2021"
}

@article{Meltzer:2020qbr,
    author = "Meltzer, David and Sivaramakrishnan, Allic",
    title = "{CFT unitarity and the AdS Cutkosky rules}",
    eprint = "2008.11730",
    archivePrefix = "arXiv",
    primaryClass = "hep-th",
    reportNumber = "CALT-TH-2020-032",
    doi = "10.1007/JHEP11(2020)073",
    journal = "JHEP",
    volume = "11",
    pages = "073",
    year = "2020"
}

@article{Maldacena:2011nz,
    author = "Maldacena, Juan M. and Pimentel, Guilherme L.",
    title = "{On graviton non-Gaussianities during inflation}",
    eprint = "1104.2846",
    archivePrefix = "arXiv",
    primaryClass = "hep-th",
    reportNumber = "PUPT-2371",
    doi = "10.1007/JHEP09(2011)045",
    journal = "JHEP",
    volume = "09",
    pages = "045",
    year = "2011"
}

@article{Chowdhury:2024SD,
    author = "Chowdhury, Chandramouli and Doran, George and Lipstein, Arthur and Monteiro, Ricardo and Nagy, Silvia and Singh, Kajal",
    title = "{Light-cone actions and correlators of self-dual theories in AdS$_{4}$}",
    eprint = "2411.04172",
    archivePrefix = "arXiv",
    primaryClass = "hep-th",
    reportNumber = "QMUL-PH-24-24",
    doi = "10.1007/JHEP01(2025)172",
    journal = "JHEP",
    volume = "01",
    pages = "172",
    year = "2025"
}

@article{Donath:2024utn,
    author = "Donath, Yaniv and Pajer, Enrico",
    title = "{The in-out formalism for in-in correlators}",
    eprint = "2402.05999",
    archivePrefix = "arXiv",
    primaryClass = "hep-th",
    doi = "10.1007/JHEP07(2024)064",
    journal = "JHEP",
    volume = "07",
    pages = "064",
    year = "2024"
}

@article{Armstrong:2023phb,
    author = "Armstrong, C. and Goodhew, H. and Lipstein, A. and Mei, J.",
    title = "{Graviton trispectrum from gluons}",
    eprint = "2304.07206",
    archivePrefix = "arXiv",
    primaryClass = "hep-th",
    doi = "10.1007/JHEP08(2023)206",
    journal = "JHEP",
    volume = "08",
    pages = "206",
    year = "2023"
}

@article{Ghosh:2014kba,
    author = "Ghosh, Archisman and Kundu, Nilay and Raju, Suvrat and Trivedi, Sandip P.",
    title = "{Conformal Invariance and the Four Point Scalar Correlator in Slow-Roll Inflation}",
    eprint = "1401.1426",
    archivePrefix = "arXiv",
    primaryClass = "hep-th",
    reportNumber = "ICTS-2013-23, TIFR-TH-13-31",
    doi = "10.1007/JHEP07(2014)011",
    journal = "JHEP",
    volume = "07",
    pages = "011",
    year = "2014"
}

@article{McFadden:2009fg,
    author = "McFadden, Paul and Skenderis, Kostas",
    title = "{Holography for Cosmology}",
    eprint = "0907.5542",
    archivePrefix = "arXiv",
    primaryClass = "hep-th",
    reportNumber = "ITF-22",
    doi = "10.1103/PhysRevD.81.021301",
    journal = "Phys. Rev. D",
    volume = "81",
    pages = "021301",
    year = "2010"
}

@article{Cespedes:2025dnq,
    author = "Cespedes, Sebastian and Jazayeri, Sadra",
    title = "{The massive flat space limit of cosmological correlators}",
    eprint = "2501.02119",
    archivePrefix = "arXiv",
    primaryClass = "hep-th",
    doi = "10.1007/JHEP07(2025)032",
    journal = "JHEP",
    volume = "07",
    pages = "032",
    year = "2025"
}

@article{Senatore:2009cf,
    author = "Senatore, Leonardo and Zaldarriaga, Matias",
    title = "{On Loops in Inflation}",
    eprint = "0912.2734",
    archivePrefix = "arXiv",
    primaryClass = "hep-th",
    doi = "10.1007/JHEP12(2010)008",
    journal = "JHEP",
    volume = "12",
    pages = "008",
    year = "2010"
}

@article{Farrow:2018yni,
    author = "Farrow, Joseph A. and Lipstein, Arthur E. and McFadden, Paul",
    title = "{Double copy structure of CFT correlators}",
    eprint = "1812.11129",
    archivePrefix = "arXiv",
    primaryClass = "hep-th",
    doi = "10.1007/JHEP02(2019)130",
    journal = "JHEP",
    volume = "02",
    pages = "130",
    year = "2019"
}

@article{Sleight:2021plv,
    author = "Sleight, Charlotte and Taronna, Massimo",
    title = "{From dS to AdS and back}",
    eprint = "2109.02725",
    archivePrefix = "arXiv",
    primaryClass = "hep-th",
    doi = "10.1007/JHEP12(2021)074",
    journal = "JHEP",
    volume = "12",
    pages = "074",
    year = "2021"
}

@article{MdAbhishek:2025dhx,
    author = "Abhishek, Md. and Sleight, Charlotte and Taronna, Massimo",
    title = "{Cosmological Correlators in Gauge Theory and Gravity from EAdS}",
    eprint = "2509.09536",
    archivePrefix = "arXiv",
    primaryClass = "hep-th",
    month = "9",
    year = "2025"
}

@article{Heckelbacher:2022hbq,
    author = "Heckelbacher, Till and Sachs, Ivo and Skvortsov, Evgeny and Vanhove, Pierre",
    title = "{Analytical evaluation of cosmological correlation functions}",
    eprint = "2204.07217",
    archivePrefix = "arXiv",
    primaryClass = "hep-th",
    reportNumber = "IPhT-t22/02, LMU-ASC 13/22",
    doi = "10.1007/JHEP08(2022)139",
    journal = "JHEP",
    volume = "08",
    pages = "139",
    year = "2022"
}

@article{Veltman:1963th,
    author = "Veltman, M. J. G.",
    title = "{Unitarity and causality in a renormalizable field theory with unstable particles}",
    doi = "10.1016/S0031-8914(63)80277-3",
    journal = "Physica",
    volume = "29",
    pages = "186--207",
    year = "1963"
}

@article{tHooft:1973wag,
    author = "'t Hooft, Gerard and Veltman, M. J. G.",
    title = "{DIAGRAMMAR}",
    reportNumber = "CERN-73-09",
    doi = "10.1007/978-1-4684-2826-1_5",
    journal = "NATO Sci. Ser. B",
    volume = "4",
    pages = "177--322",
    year = "1974"
}

@article{Werth:2024mjg,
    author = "Werth, Denis",
    title = "{Spectral representation of cosmological correlators}",
    eprint = "2409.02072",
    archivePrefix = "arXiv",
    primaryClass = "hep-th",
    doi = "10.1007/JHEP12(2024)017",
    journal = "JHEP",
    volume = "12",
    pages = "017",
    year = "2024"
}

@article{Meltzer:2021bmb,
    author = "Meltzer, David",
    title = "{Dispersion Formulas in QFTs, CFTs, and Holography}",
    eprint = "2103.15839",
    archivePrefix = "arXiv",
    primaryClass = "hep-th",
    reportNumber = "CALT-TH-2021-012",
    doi = "10.1007/JHEP05(2021)098",
    journal = "JHEP",
    volume = "05",
    pages = "098",
    year = "2021"
}

@article{Meltzer:2021zin,
    author = "Meltzer, David",
    title = "{The inflationary wavefunction from analyticity and factorization}",
    eprint = "2107.10266",
    archivePrefix = "arXiv",
    primaryClass = "hep-th",
    reportNumber = "CALT-TH-2021-028",
    doi = "10.1088/1475-7516/2021/12/018",
    journal = "JCAP",
    volume = "12",
    number = "12",
    pages = "018",
    year = "2021"
}

@article{Das:2025qsh,
    author = "Das, Shibam and Karan, Debanjan and Khatun, Babli and Kundu, Nilay",
    title = "{A single-cut discontinuity for cosmological correlators from unitarity and analyticity}",
    eprint = "2512.20720",
    archivePrefix = "arXiv",
    primaryClass = "hep-th",
    month = "12",
    year = "2025"
}

@article{Ansari:2026xkm,
    author = "Ansari, Arhum and Jain, Sachin and Mazumdar, Deep",
    title = "{Cosmological Cutting Rules from Flat-Space Unitarity via Dressing}",
    eprint = "2601.08917",
    archivePrefix = "arXiv",
    primaryClass = "hep-th",
    month = "1",
    year = "2026"
}

@article{Tong:2021wai,
    author = "Tong, Xi and Wang, Yi and Zhu, Yuhang",
    title = "{Cutting rule for cosmological collider signals: a bulk evolution perspective}",
    eprint = "2112.03448",
    archivePrefix = "arXiv",
    primaryClass = "hep-th",
    doi = "10.1007/JHEP03(2022)181",
    journal = "JHEP",
    volume = "03",
    pages = "181",
    year = "2022"
}

@article{Ema:2024hkj,
    author = "Ema, Yohei and Mukaida, Kyohei",
    title = "{Cutting rule for in-in correlators and cosmological collider}",
    eprint = "2409.07521",
    archivePrefix = "arXiv",
    primaryClass = "hep-th",
    reportNumber = "UMN-TH-4331/24, FTPI-MINN-24-18, KEK-TH-2653",
    doi = "10.1007/JHEP12(2024)194",
    journal = "JHEP",
    volume = "12",
    pages = "194",
    year = "2024"
}

@article{Colipi-Marchant:2025oin,
    author = "Colip{\'\i}-Marchant, Francisco and Marin, Gabriel and Palma, Gonzalo A. and Rojas, Francisco",
    title = "{Schwinger-Keldysh Cosmological Cutting Rules}",
    eprint = "2512.22652",
    archivePrefix = "arXiv",
    primaryClass = "hep-th",
    month = "12",
    year = "2025"
}

@article{abreu:2015cuts,
  title={Cuts and coproducts of massive triangle diagrams},
  author={Abreu, Samuel and Britto, Ruth and Gr{\"o}nqvist, Hanna},
  journal={Journal of High Energy Physics},
  volume={2015},
  number={7},
  pages={1--60},
  year={2015},
  publisher={Springer}
}

@article{Chavez:2012kn,
    author = "Chavez, Federico and Duhr, Claude",
    title = "{Three-mass triangle integrals and single-valued polylogarithms}",
    eprint = "1209.2722",
    archivePrefix = "arXiv",
    primaryClass = "hep-ph",
    doi = "10.1007/JHEP11(2012)114",
    journal = "JHEP",
    volume = "11",
    pages = "114",
    year = "2012"
}

@article{Abreu:2014cla,
    author = "Abreu, Samuel and Britto, Ruth and Duhr, Claude and Gardi, Einan",
    title = "{From multiple unitarity cuts to the coproduct of Feynman integrals}",
    eprint = "1401.3546",
    archivePrefix = "arXiv",
    primaryClass = "hep-th",
    doi = "10.1007/JHEP10(2014)125",
    journal = "JHEP",
    volume = "10",
    pages = "125",
    year = "2014"
}

@article{Goodhew:2024eup,
    author = "Goodhew, Harry and Thavanesan, Ayngaran and Wall, Aron C.",
    title = "{The Cosmological CPT Theorem}",
    eprint = "2408.17406",
    archivePrefix = "arXiv",
    primaryClass = "hep-th",
    month = "8",
    year = "2024"
}

@article{Rajaraman:2010xd,
    author = "Rajaraman, Arvind",
    title = "{On the proper treatment of massless fields in Euclidean de Sitter space}",
    eprint = "1008.1271",
    archivePrefix = "arXiv",
    primaryClass = "hep-th",
    reportNumber = "UCI-TR-2010-14",
    doi = "10.1103/PhysRevD.82.123522",
    journal = "Phys. Rev. D",
    volume = "82",
    pages = "123522",
    year = "2010"
}

@article{Beneke:2012kn,
    author = "Beneke, M. and Moch, P.",
    title = "{On {\textquotedblleft}dynamical mass{\textquotedblright} generation in Euclidean de Sitter space}",
    eprint = "1212.3058",
    archivePrefix = "arXiv",
    primaryClass = "hep-th",
    reportNumber = "TUM-HEP-870-12, TTK-12-49",
    doi = "10.1103/PhysRevD.87.064018",
    journal = "Phys. Rev. D",
    volume = "87",
    pages = "064018",
    year = "2013"
}

\end{document}